\documentclass[preprint,12pt]{elsarticle}

\usepackage{color}
\usepackage{amsmath}
\usepackage{hyperref}
\usepackage{amssymb}
\usepackage[dvipsnames]{xcolor}
\usepackage{capt-of}

\newcommand{\argmax}{\mathop{\mathrm{argmax}}}

\journal{Mathematical Biosciences}

\begin{document}

\begin{frontmatter}



\title{Agent-based Dynamics of a SPAHR Opioid Model on Social Network Structures}


\author[utk]{Owen Queen}

\affiliation[utk]{organization={Department of Mathematics},
            addressline={University of Tennessee, Knoxville}, 
            city={Knoxville},
            postcode={37996-1320}, 
            state={TN},
            country={USA}}

\author[utk]{Vincent Jodoin}
\author[utk]{Leigh B. Pearcy}
\author[utk,utk-eeb]{W. Christopher Strickland\corref{cor1}}
\cortext[cor1]{Corresponding author}
\ead{cstric12@utk.edu}

\affiliation[utk-eeb]{organization={Department of Ecology and Evolutionary Biology},
            addressline={University of Tennessee, Knoxville}, 
            city={Knoxville},
            postcode={37996-1320}, 
            state={TN},
            country={USA}}

\begin{abstract}
Addiction epidemiology has been an active area of mathematical research in recent years. However, the social and mental processes involved in substance use disorders versus contraction of a pathogenic disease have presented challenges to advancing the epidemiological theory of substance abuse, especially within the context of the opioids where both prescriptions and social contagion have played a major role. In this paper, we utilize an agent-based modeling approach on social networks to further explore these dynamics. Using parameter estimation approaches, we compare our results to that of the Phillips \textit{et al.} SPAHR model which was previously fit to data from the state of Tennessee. Our results show that the average path length of a social network has a strong relationship to social contagion dynamics for drug use initiation, while other pathways to substance use disorder should not be constrained to social network interactions that predate the individual's drug use.
\end{abstract}




\end{frontmatter}


\section{Introduction}
\label{sec:Introduction}

Substance use disorder and overdose mortality related to opioid use continues to be a major public health issue in the United States \cite{CDC2021-basics}. There are signs that the COVID-19 epidemic has only exacerbated the problem, especially with regard to risk of overdose death \cite{Slavova2020, CDC2021-fentanyl}. Illicitly manufactured fentanyl, a synthetic opioid that is up to 50 times more potent than heroin and up to 100 times stronger than morphine, is behind much of this mortality. Some recent fentanyl analogs are estimated to be even stronger - up to 10,000 times more potent than morphine in the case of Carfentanil. These are often found mixed with other drugs, particularly heroin, cocaine, and methamphetamine, and made into pills which resemble other prescription opioids. Taken together, it is estimated that these non-methadone synthetic opioids account for 73\% of all opioid-involved deaths and are the most common drugs involved in overdose deaths of any sort \cite{CDC2021-fentanyl,CDC2021-stopoverdose,Phillips2021,Wilson2020}.

There have been numerous simulation and conceptual modeling studies targeting some form of opioid misuse despite a noted lack of financial support for modeling studies on the opioid crisis from public health organizations \cite{Maglio2014,Sharareh2019,Cerda2021}. Among simulation models, the most frequent approach is compartmental modeling, followed by Markov models, system dynamics models, and agent-based models. However, a recent review of opioid simulation models found that fewer than half presented model equations or provided access to model code and documentation, making it impossible to adequately interpret the findings, reproduce the results, or meaningfully establish how differences in mechanistic structure can lead to different qualitative conclusions \cite{Cerda2021}. 

While careful, well-justified mechanistic development and overall transparency is important in any modeling study, there are also trade-offs related to model complexity. This is particularly true for agent-based models (ABMs). Complex, detailed ABMs offer a high degree of realism that is attractive to policy makers and can provide a virtual laboratory for testing management strategies. On the other hand, they can be challenging to parameterize from data and lose generality and tractability to structural analysis that can be critical for advancing theory \cite{Levins1966,Reardon2018}. If the goal is to inform theory and advance mathematical results relating to model structure, parsimony is critical, and minimalist ABMs can be powerful tools for forging a connection between key, individual-level behaviors and population-level phenomena. This understanding can then be used to form mathematical models of population dynamics based on the individual-level mechanisms.

Our study takes this reductionist approach to agent-based modeling in order to study the role of social network structure on the theoretical results of an ordinary differential equation (ODE) compartmental model for opioid use disorder. The current US opioid epidemic is driven by a combination of prescribing practices and social factors \cite{Battista2019,Cutter2021,Phillips2021}. Incorporating both of these mechanisms into an ODE compartmental model results in a different model structure than typically seen in infectious disease and exclusively socially-driven drug use disorder settings, with the result that typical approaches for analysis relying on $\mathcal{R}_0$ no longer apply \cite{Battista2019,Phillips2021}. The effect of social networks in this mathematical setting has yet to be explored, but there is plenty of evidence that it plays a key role. Multiple studies reveal a connection between friend and family opioid use and opioid use initiation, and there are  strong arguments in favor of applying social contagion theory to opioid use disorder \cite{SAMHSA2016,Seamans2018,Harbaugh2019,Cutter2021}.

Given the importance of prescription opioids and fentanyl to the current state of the US opioid epidemic, our social-network ABM study is based on a recent, data-driven ODE model which focuses on the interconnected dynamics of both of these factors. This Phillips \textit{et al.} model, like many ODE models, inherently assumes the well mixing principle. A population is described as well-mixed if every individual in the population interacts with all of the others, but models often assume that the well-mixing principle reasonably describes phenomena in populations that may only be approximately well mixed. However, these models may break down upon consideration of populations whose social network structure significantly deviates from well-mixed \cite{DURRETT1994363}. In this study, we will seek to determine the relative influence of social network based contagion on the spread of illicit- and prescription-sourced opioids \cite{Cutter2021}.

The remainder of this paper is organized as follows. First, we briefly describe the Phillips \textit{et al.} model and our mathematical approach to considering it as the mean-field, population-level model for an individual-level stochastic process. Next, we describe the construction of several social network models that are used for comparison purposes in our methods, including how vertices are to be removed and added in during the course of a model simulation. We then describe our procedure for comparing ABM parameterization to that of the Phillips \textit{et al.} ODE model. Results are presented showing how different social network metrics are related to substance use disorder outcomes, with average path length showing the strongest relationship. We then use a parameter estimation procedure to show that prescription opioid based heroin and fentanyl initiation must be independent of the social network in order to reproduce the results of Phillips \textit{et al.}, with direct rates of substance use disorder (from the $S$ class directly into $A$ and $H$) left to adjust for the model's social network structure. Finally, we relate average path length directly to the value of the $S$ to $H$ rate, suggesting that social network dynamics almost exclusively affect social contagion dynamics whereby susceptible individuals acquire a heroin or fentanyl use disorder directly, and not through the use of prescription opioids first.

\section{Methods}
\label{sec:Methods}

The agent-based model (ABM) developed in this study is based on a study from Phillips \textit{et al.} \cite{Phillips2021} which described a five-class, SPAHR model for prescription- and illicit-based opioid use disorder. Their model is formulated as a system of ordinary differential equations (ODEs) and serves as the underlying model for our extensions here; therefore, we shall often refer to the Phillips model as the ``ODE model'' versus our ABM-based, stochastic work. A consequence of this is that many of the modeling assumptions used in this project are inherited from the Phillips \textit{et al.} ODE model. We refer the reader to the Phillips \textit{et al.} study \cite{Phillips2021} for a complete discussion of these assumptions and their consequences. 

\begin{figure} 
    \centering
    \includegraphics[width=\linewidth]{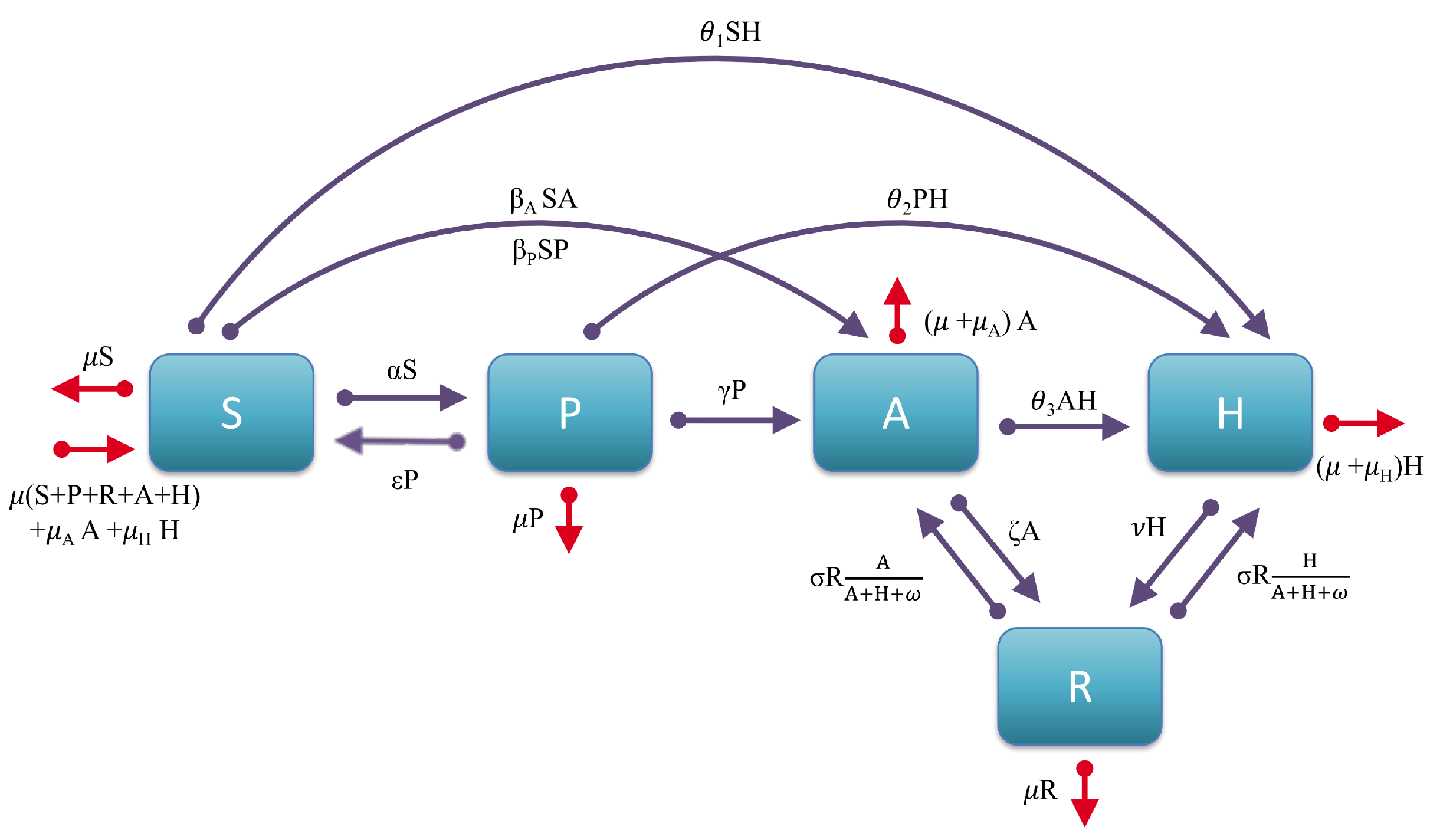}
    \caption{Compartment diagram from Phillips \textit{et al.} \cite{Phillips2021}. Both the definition of the compartments and their basic relationships remain the same in our ABM.}
    \label{fig:schematic_diagram}
\end{figure}

Both the ABM and the ODE model contain the same compartments representing the state of individuals in a given population. These can be described as follows:
\begin{enumerate}
    \item Susceptible (S): Individuals are not taking prescription opioids or heroin or fentanyl, and they have not previously suffered from opioid use disorder. 
    \item Prescribed (P): Individuals are taking prescription opioids, but their use patterns do not qualify as a disorder. 
    \item Addiction to prescription opioids (A): Individuals have a use disorder related to prescription opioids, but they are not using heroin or fentanyl. 
    \item Heroin addiction (H): Individuals have an opioid use disorder which includes heroin or fentanyl. It may still also include prescription opioids.
    \item Stably recovered (R): Individuals who quit taking opioids and/or complete treatment for opioid use disorder and do not relapse within 4 weeks.
\end{enumerate}
Phillips \textit{et al.} \cite{Phillips2021} considered these compartments as proportions of the entire population so that $S+P+A+H+R=1$. In the case of our ABM, we will assume that these classes are both mutually exclusive and exhaustive, with each agent belonging to only one of the classes at any given time step. The Phillips \textit{et al.} ODE model can be expressed as
\footnotesize
\begin{align}
\frac{dS}{dt} & = -\alpha(t) S - \beta_{A} SA  -\beta_{P} SP- \theta_{1} SH +\varepsilon P + \mu (P+A+H+R)\nonumber\\
&\ \ \ \ + \mu_{A}(t)A + \mu_{H} H \nonumber \\
\frac{dP}{dt} & = \alpha S - \varepsilon P  - \gamma P - \theta_{2}PH- \mu P \nonumber  \\
\label{eq:ODEs} \frac{dA}{dt} & = \gamma P + \sigma R \frac{A}{A+H+\omega} +\beta_{A} SA  +\beta_{P} SP -\zeta A - \theta_{3}AH-(\mu + \mu_{A}(t))A \\
\frac{dH}{dt} & = \theta_{1}SH+\theta_{2}PH+\theta_{3}AH + \sigma R \frac{H}{A+H+\omega}-\nu H-(\mu+\mu_{H})H \nonumber \\
\frac{dR}{dt} & = \zeta A +\nu H -\sigma R \frac{A}{A+H+\omega}-\sigma R \frac{H}{A+H+\omega} -\mu R \nonumber \\
\text{with} \nonumber \\
1 & =S+P+A+H+R. \nonumber 
\end{align}
\normalsize
A schematic of this model from Phillips \textit{et al.} \cite{Phillips2021} is shown in Fig. \ref{fig:schematic_diagram}. In addition to the compartments listed above, all of the ODE transition pathways will be represented in our ABM. The Phillips model consists of validated parameter values based on data from the state of Tennessee, so we will leverage these values and the model's time series outputs for parameterization and evaluation of our ABM results. However, the time-varying parameters from the Phillips \textit{et al.} model will be set at their initial values and made constant for the purpose of our study, both for simplicity and to focus on the network effects of the ABM versus the ODE model formulation.

\subsection{Converting the ODE model to an agent-based model}

Given the work conducted by Phillips \textit{et al.} on the deterministic, mean-field model represented by System \ref{eq:ODEs}, switching to an agent-based formulation conveys certain advantages for further analysis. A primary benefit is the individual-level characterization of agent-to-agent interaction versus the population-level, mass-action formulation of the ODEs. Using an individual-level approach, it becomes natural to explicitly model preexisting social connections between individuals with a network and then directly consider both different network structures and different scenarios for agent-to-agent interaction that could depend on that network. We can also begin to quantify a certain degree of uncertainty due to individual-level effects by bootstrapping the results of our simulations, though this comes with a computational cost versus the deterministic, ODE approach.

Our process for converting the Phillips \textit{et al.} ODE model to an agent-based model (ABM) proceeds as follows. In the ODE model, we assume that each compartment represents the mean expected fraction of a total population which belongs to that class, and that each term in the system of equations defines a mean rate of change between these compartments. In the case of a linear term, e.g. $\alpha S$, the parameter ($\alpha$) then defines a constant, mean rate of transition (in this case, from $S$ to $P$) per individual in the $S$ compartment per unit time (years for the Phillips \textit{et al.} model \cite{Phillips2021} and in our study). These transitions also occur independently of the time since the last event, as all information about individuals or past events in the ODE model are lost. Multiplying by the relevant compartment, which is always the compartment making the transition ($S$ in our $\alpha S$ example), then gives the total expected number of individuals (or the population fraction in the units of Phillips \textit{et al.} \cite{Phillips2021}) that make the transition. By definition, this also implies that individuals in the model are undergoing a Poisson process with rate parameter $\alpha$, and we can therefore model the waiting time before an individual's transition between compartments with an exponential distribution.

In the case of the ODE model's nonlinear rates, this line of reasoning only changes slightly. As with the linear rates, each nonlinear term of the ODE model always includes the class that is making the transition. Setting aside the relapse rates momentarily, each of the other nonlinear rates also includes a second class which the transitioning class must interact with in some way, whether directly (e.g. via social contact) or indirectly (e.g. drug availability as a function of current demand, see Phillips \textit{et al.} \cite{Phillips2021} for details). We can optionally relax the well-mixing assumption of the ODE model by assuming that these interactions take place according to a social network with average ``infection'' rate given by the coefficient of the corresponding ODE rate term. 

For our study, we will assume that this network is an undirected, simple graph where the nodes are agents and the edges represent significant social interaction between the agents. The network will be fixed at the start of each simulation and will not change except to accommodate new agents which come into the network to replace a departing agent that underwent a death process during a time-step. This process is specific to the network generation algorithm chosen and will be explained in more detail later.

\begin{figure}[t]
    \centering
    \includegraphics[width=\textwidth]{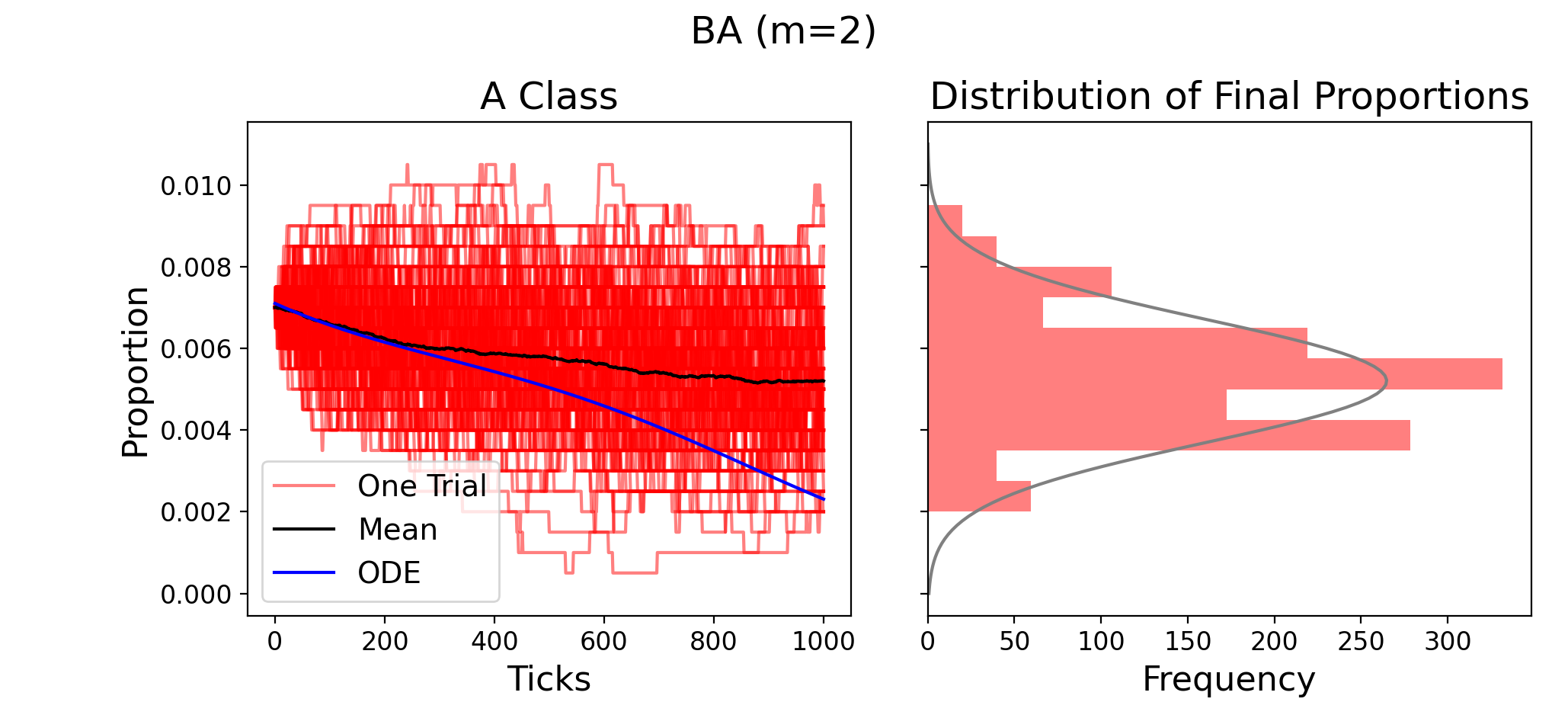}
    \caption{Example of bootstrapping the $A$ class in the SPAHR model using our ABM with 200 trials. Each single trial is a red trajectory on the left (1 model tick = 0.01 years) and the point-wise mean is shown in black. The distribution of ending proportions is shown on the left. As expected, this distributions roughly resembles a Gaussian distribution (shown in grey). The Barab\'{a}si-Albert model with parameter value $m=2$ was used in this example. Note that there is strong discrepancy between the mean of the ABM and the ODE trajectory (shown in blue). This is expected due to the deviation in social network structure introduced by the Barab\'{a}si-Albert model.}
    \label{fig:bootstrap}
\end{figure}

In the case of a social interaction via the network, the relative exposure of an individual agent to a substance using class, for example, the $H$-class, is given by the number of its network neighbors in $H$ divided by its total number of neighbors. In the case of the relapse rates in the ODE model, the quotient terms are meant to determine the class into which individuals relapse. Since it is quite possible an individual in $R$ has no neighbors in either $A$ or $H$, the global quotient (using the total number of $A$ and $H$ in the entire network) is used instead for this transition.

In all cases, our agent-based model uses a two-step approach to determining agent transitions between classes. First, for a given agent with a given class attribute, all mean transition rates out of the current class (as given by the ODE model) are added together into a parameter $\lambda$. Under the assumption that this sum defines a mean waiting time in a Poisson process, we model the probability that the agent transitions out of its class in a time-step $\Delta t$ to be
\[
P\big( \textrm{transition in } (t, t + \Delta t] \big) = 1 - e^{-\lambda \Delta t}
\]
where
\[
\lambda = \sum a_i + \sum b_j N_j.
\]
$a_i$ denotes all linear coefficients of rates out of the current class, $b_j$ denotes all nonlinear coefficients out of the current class, and $N_j$ is the density of all neighbors in the contact class. Relapse is considered as a linear rate for this purpose, with $\sigma$ the coefficient. 

During a simulation, comparing this probability to a uniform, psuedorandom number establishes whether or not a given agent will make a transition during a time-step of length $\Delta t$. Making more than one transition per time-step is a higher order transition probability ($O(\Delta t^2)$) and is neglected in our model.

Assuming the agent will make a transition, the second step of the algorithm is to determine which class the agent transitions to among the various possibilities as determined by the directed connections in Fig. \ref{fig:schematic_diagram}. Each individual transition rate contributing to $\lambda$ is normalized by $\lambda$ and treated as a probability. If relapse is an option, $\sigma$ is divided up into $A/(A+H)$ and $H/(A+H)$ components, where $A$ and $H$ are the total number of prescription opioid addiction-class agents and heroin addiction-class agents in the simulation, and these two components define transitions probabilities into the $A$ and $H$ classes respectively. If both $A$ and $H$ are zero, we divide the probability $\sigma/\lambda$ evenly between the two transition cases. 

Lastly, since $H=0$ defines an absorbing state for the model (as long as $A>0$ or $R=0$), our model artificially converts one random $S$ agent into an $H$ whenever $H=0$. This avoids a discrepancy with the original ODE model due to discretization: if $H(0)>0$ in the ODE system, $H$ can asymptotically approach zero but never reach it. However, in a discrete, stochastic, agent-based model, it is quite possible to achieve $H=0$ for nonzero initial conditions, especially when the initial count of $H$ is relatively small. $S$ was chosen as the reservoir class for this conversion because it tends to be the class with the greatest number of individuals by far when running simulations with parameters from Phillips \textit{et al.} \cite{Phillips2021}.

The total population of the model is determined prior to simulation and is constant between time steps. When agents undergo a death process in the model, we immediately introduce another agent and assign it the $S$ class.

\subsection{Social Networks}
\label{sec:socialnetworks}


As mentioned before, preexisting social connections between discrete agents are modeled as network. We define a network, which is a type of graph, to be $G = (V, E)$, where the set $V = \{v_1, v_2, ..., v_N\}$ contains the \textit{nodes} (agents) of the network and the set $E \subseteq \{ (v_i, v_j) | \forall v_i, v_j \in V,\ i\neq j \}$ contains the \textit{edges} (preexisting social connections) of the network. Edges $(v_i, v_j)$ are considered as unordered pairs, meaning that $(v_i, v_j) = (v_j, v_i)$. 

For a network $G = (V, E)$, if there exists an undirected edge $(v_i, v_j) \in E$ then we say that $v_j$ is a \textit{neighbor} of node $v_i$ and vice versa. 
The set of all neighbors of a node $v_i$ is referred to as the \textit{neighborhood of $v_i$}, $\mathcal{N}(v_i) = \{ v_j | \forall (v_i, v_j) \in E \}$. The \textit{degree} of a node is its number of neighbors and can be thought of as a function, $\textrm{deg}:V \rightarrow \mathbb{R}$, defined as $\textrm{deg}(v_i) = |\mathcal{N}(v_i)|$ for $v_i \in V$. 

Edges in our network represent social connections, meaning any relationship where there exists a potential to spread a behavioral practice or induce a behavioral change, a phenomenon that has been called a “social contagion” \cite{hill2010infectious}. We consider each connection to be a homogeneous social interaction; that is, the strength of social interactions is considered to be equal for any pair of individuals in the network who are connected by an edge. By this definition, we can consider the neighborhood $\mathcal{N}(v_i)$ to be the epidemiologically relevant acquaintances of the agent $v_i$

In terms of network structure, we can envision a well-mixed population as a fully connected or complete graph. Therefore, we expect the mean of a large set of ABM realizations conducted on a fully connected network to approximate the Phillips \textit{et al.} model \cite{Phillips2021}. Any other network structure could potentially yield different results, though it only affects the model through the five rates described in Tbl. \ref{tab:network_rates}.
\begin{table}[t]
\centering
\begin{tabular}{ |p{1.5cm}||p{11.5cm}|  }
 \hline
 Term & Definition \\
 \hline \hline
 $\beta_A SA$   & Rate at which $S$ individuals become addicted to prescription opioids primarily by illicit purchases or interaction with $A$ individuals. \newline
 ODE: $A$ = Proportion of model population in class $A$  \newline
 ABM: $A$ = Proportion of agent’s neighbors in class $A$\\
 \hline
 $\beta_P SP$ & Rate at which $S$ individuals become addicted to prescription opioids primarily by using left-over or stolen prescription drugs. \newline
ODE: $P$ = Proportion of model population in class $P$ \newline
ABM: $P$ = Proportion of agent’s neighbors in class $P$
 \\
 \hline
 $\theta_1 SH$ & Rate at which $S$ individuals become addicted to heroin. \newline
ODE: $H$ = Proportion of model population in class $H$ \newline
ABM: $H$ = Proportion of agent’s neighbors in class $H$
 \\
 \hline
 $\theta_2 PH$ & Rate at which $P$ individuals become addicted to heroin. \newline
ODE: $H$ = Proportion of model population in class $H$ \newline
ABM: $H$ = Proportion of model population in class $H$ (no change)
\\
 \hline
 $\theta_3 AH$ & Rate at which $A$ individuals become addicted to heroin. \newline
ODE: $H$ = Proportion of model population in class $H$ \newline
ABM: $H$ = Proportion of model population in class $H$ (no change)
 \\
 \hline
\end{tabular}
\caption{Description of network-dependent rates in the ABM. Note that $\beta_A$, $\beta_P$, $\theta_1$, $\theta_2$, and $\theta_3$ are constant scalars.
}
\label{tab:network_rates}
\end{table}

In addition to the fully connected network, we examined the effect of networks created from three common network generation algorithms: the Erd\H{o}s-R\'{e}nyi network \cite{ER_network}, the Barab\'{a}si-Albert network \cite{BA_network}, and the Watts-Strogatz network \cite{WS_network}. In any given ABM simulation, one of these algorithms is specified (or the complete network) and the network is generated for the requested number of agents. 

However, when agents die in the model, we felt it overly artificial to reintroduce a susceptible agent with the same connections as before. Instead, if $n$ individuals die in time step $t$, then those $n$ individuals are removed entirely from the network, including any edges associated with those individuals. Then $n$ new individuals are immediately introduced into the network, forming new connections and each being assigned class $S$. In order to preserve the properties of the original network generation algorithm as closely as possible, we implemented network generation-specific reintroduction algorithms that define how a new node is introduced into the network after the removal of an old node. Descriptions of each network generation algorithm and their corresponding reintroduction algorithms are given below.


\subsubsection{Erd\H{o}s-R\'{e}nyi Model}
The Erd\H{o}s-R\'{e}nyi random graph has served as a baseline model for constructing random networks \cite{ER_network}. In this network generation algorithm, $n$ nodes are created, and every edge $(v_i, v_j)$ between two nodes $v_i$ and $v_j$ has an equal probability $p$ of being included in the network.

When a new agent (node) $v_{n+1}$ is introduced into the network during a simulation, each possible edge with the new node, $\{(v_i, v_{n+1})|\forall i \in (1, 2, ..., n)\}$, is added with the same probability $p$ as used in the original network generation.

\subsubsection{Barab\'{a}si-Albert Model}
The Barab\'{a}si-Albert network \cite{BA_network} is a scale-free network generation algorithm that focuses on preferential attachment. Nodes are added in sequence to generate a graph, and for each newly-added node, the probability $\Pi$ of forming an edge connecting the new node with any other node $v_i$ depends on the degree of $v_i$, $\textrm{deg}(v_i)$ via

\[\Pi(v_i) = \frac{\textrm{deg}(v_i)}{\sum_j\textrm{deg}(v_j)}\]

This network generation algorithm has been extensively studied, and the preferential attachment and resulting scale-free degree distribution have a basis in observations of real-world networks \cite{barabasi2013network}. Since this network is built in a sequential manner, our reintroduction algorithm is defined in the same way: Edge connections draw on the distribution described by $\Pi$ just as if they were coming into the network at initialization.

\subsubsection{Watts-Strogatz Model}

\begin{figure} 
    \centering
    \includegraphics[width=\linewidth]{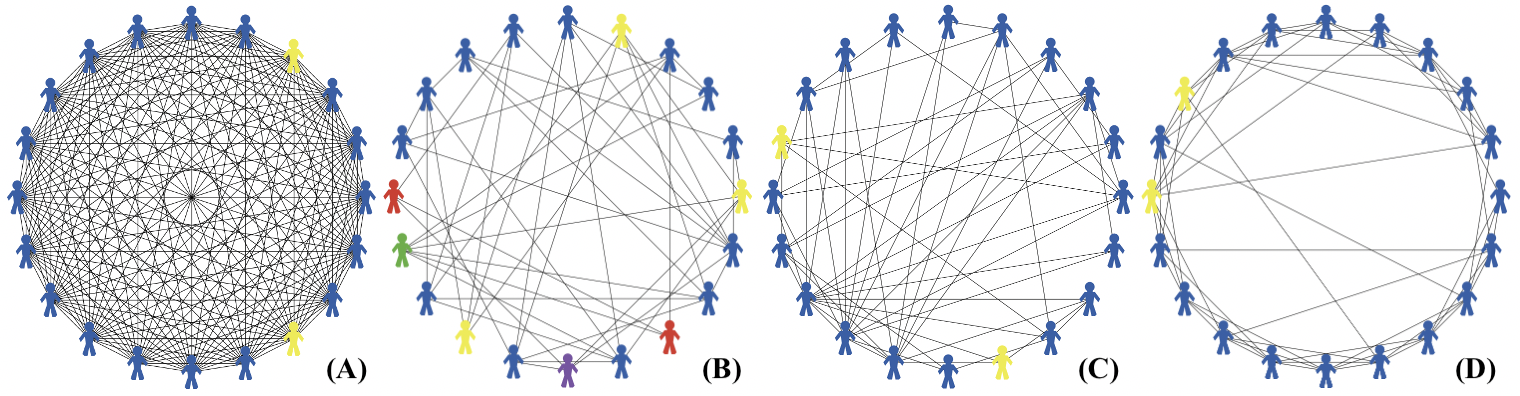}
    \caption{These plots exhibit varying social network structures utilized by the ABM. (A) is a fully connected network, a structure that was used to model the well-mixed assumption from the ODE model by Phillips \textit{et al.}; (B) shows an Erd\H{o}s-R\'{e}nyi network; (C) shows a Barab\'{a}si-Albert network; and (D) shows a Watts-Strogatz network. }
    \label{fig:message}
\end{figure}

The Watts-Strogatz network generational algorithm, also known as a small-world network, has been shown to demonstrate properties consistent with real-world social networks \cite{WS_network}. The stochasticity of this model is also more controlled than Barab\'{a}si-Albert or Erd\H{o}s-R\'{e}nyi networks, with the amount of randomness in the network being controlled by one parameter. 


The Watts-Strogatz algorithm defines three parameters for building the network $G = (V, E)$:
\begin{itemize}
    \item $N$: Total number of nodes in the network, where $|V| = N$. In the ABM, this will correspond to the total population size.
    \item $n$: Known as the ``neighborhood size." Each node is initially connected to the $n$ closest nodes in the lattice structure (a process that will be described in subsequent sections). This parameter also fixes the eventual mean degree of all nodes in the network to be $n$, i.e. $\frac{1}{N}\sum\limits_{i=1}^{N}\textrm{deg}(v_i) = n$. By the definition given in \cite{WS_network}, $n$ must be an even number. Note that in \cite{WS_network}, $n$ is synonymous to the $k$ parameter.
    \item $p$: Known as the ``rewire probability." This is the probability that any edge within the original lattice may be disconnected and reconnected to another randomly-chosen node in the network. One can think of $p$ as measuring the level of disorder in a Watts-Strogatz network, with $p=0$ resulting in a regular lattice and $p=1$ resulting in a random graph, similar to the Erd\H{o}s-R\'{e}nyi network.
\end{itemize}

In the Watts-Strogatz network generation algorithm, $N$ nodes are arranged in a lattice. An edge is then created between each node and the $n$ nodes closest to it within the lattice structure. After creating this initial structure, a Bernoulli process is then performed on the set of edges with each edge having a probability $p$ of one of the nodes on the edge being swapped for another. For more information on this algorithm, we direct the reader to \cite{WS_network}.

We are not aware of any established algorithm for the introduction of a new node into a Watts-Strogatz network, so we will outline our method here. On a high level, this Watts-Strogatz reintroduction algorithm strives to reintroduce new agents in the place of dead agents at the end of some time step $t_i$ while approximately preserving the local network topology of the Watts-Strogatz network.

Consider a Watts-Strogatz network defined by a graph $G = (V, E)$ generated with parameter values $N$, $n$, and $p$. In the description of this algorithm, we will consider the terms ``nodes" and ``agents" to be synonymous when referring to operations on the network $G$. Let $D \subseteq V$ be the subset of agents who die as a result of the ABM transitions. The first step in the algorithm is to remove all dead agents from the network, i.e., all edges belonging to any node in $D$ are removed from $E$. This leaves each dead agent with no edges connecting it to the network. In the second step of the algorithm, they will be reintroduced as new nodes with their class property set to $S$. We will assume that $|V \setminus D| > 0$ so that $D$ is a proper subset of $V$. If this is not the case, the entire network is simply reconstructed using the standard Watts-Strogatz algorithm with the same parameters.

The reintroduction algorithm can be broken down into two separate portions:
\begin{enumerate}
    \item \textbf{Pre-rewire neighborhood identification}: A stochastic process is performed to identify a set of nodes that is highly-clustered and serves as an initial neighborhood for the reintroduced node. This is analogous to the initial Watts-Strogatz network generation, where nodes are connected to nearest neighbors within the lattice before edges are rewired. 
    \item \textbf{Rewiring procedure}: Rewiring, as in the original Watts-Strogatz algorithm, is performed on the pre-rewire edge set of the reintroduced node.
\end{enumerate}

\paragraph{Pre-rewire neighborhood identification} Consider a ``dead'' node $v_d \in D$ with all edges removed. We will describe the construction of $\mathbf{A}$, the pre-rewire neighborhood of $v_d$. Since we wish to keep the average degree of all nodes in the network approximately equal to $n$, we will add $n$ new edges connecting $v_d$ to the existing network.

$\mathbf{A}$ will be built through an iterative process. Let $\mathbf{A}_m = \{ v_1, ..., v_m | m < n, v_i \in V\} \subseteq V$ be the intermediate version of $\mathbf{A}$ at some iteration $m < n$ in the process. 
Let $\mathbf{N}_m$ be the set of nodes given by the union of all neighbors of nodes $v_i \in \mathbf{A}_m$ but excluding any nodes already contained in $\mathbf{A}_m$, i.e.
\[\mathbf{N}_m = \bigcup\limits_{i=1}^{m} \mathcal{N}(v_i) \setminus \mathbf{A}_m\]
where $\mathcal{N}(v_i)$ denotes the neighborhood of $v_i$. Define a function $\textrm{deg}_{\mathbf{A}_m}: \mathbf{N}_m \rightarrow \mathbb{R}$ such that for any node $v \in \mathbf{N}_m$, $\textrm{deg}_{\mathbf{A}_m}$ is the number of edges connecting $v$ to any node in $\mathbf{A}_m$. Therefore,
\[\textrm{deg}_{\mathbf{A}_m}(v) = \Big|\{ (v, v_j) | v_j \in \mathbf{A}_m \text{ and } (v, v_j) \in E\}\Big|\]


To begin the algorithm, we choose a random node $v_1 \in V \setminus D$  and set $\mathbf{A}_1 = \{v_1\}$. At all successive iterations, we will choose the next node $v_{m+1}$ to be the one with the maximum number of edges connecting it to any nodes in the current collection $\mathbf{A}_m$, i.e.
\[v_{m+1} = \argmax\limits_{v \in \mathbf{N}_m} \ \textrm{deg}_{\mathbf{A}_m}(v)\]
If $\mathbf{N}_m$ is empty then it is first checked whether $\mathbf{A}_m = V$. If this is the case, a node from $D$ is chosen at random to be assigned to $v_{m+1}$ (this does not prevent the reintroduction algorithm from acting on this node in the future if it hasn't already). If not, the network is not connected, and a node is chosen from $V \setminus \mathbf{A}_m$.


If there is more than one choice for this $v_{m+1}$, we take one at random from the candidates. Adding this new node to $\mathbf{A}_m$ gives us $\mathbf{A}_{m+1}$. Repeating this process for $n$ iterations gives us the full set $\mathbf{A}$.

\paragraph{Rewiring procedure} Once $\mathbf{A}$ is obtained for a given node $v_d$, we then perform rewiring in an identical procedure to the one described in the original Watts-Strogatz algorithm but restricted to edges in the set $\{ (v_d, v_i) | v_i \in \mathbf{A} \}$. That is, each edge connected to $v_d$ has a probability $p$ of being rewired to a random node in $V$ where $p$ is the parameter used to generate the starting Watts-Strogatz algorithm network. This procedure concludes the reintroduction of $v_d$ into the network.

This reintroduction algorithm is repeated for all $v_d \in D$. Additionally, we repeat the reintroduction procedure for every disconnected agent at each time step $t_i$ in the simulation so that there are never any isolated nodes, but the class property of these agents is preserved rather than reset to $S$. This reintroduction prevents disconnected agents from persisting within the network during a time-step, a phenomenon that would disrupt the network-based rates in the ABM.

\subsection{ABM parameter fitting}
\label{sec:opt_methods}

For parameters that were not network-dependent, we used estimated values from Phillips \textit{et al.} derived from Tennessee data \cite{Phillips2021}. In Phillips \textit{et al.} \cite{Phillips2021}, $\alpha$ and $\mu_a$ were defined as time-varying parameters. However, in our work, we keep these parameters constant at their initial values in order to better focus on comparing the ODE model results, which represent a fully connected community, to the various social network structures used in the ABM.

In the context of the data-fitted ODE model of Phillips \textit{et al.}, a primary question for our study was whether or not the output of their fitted model could be reproduced with an agent-based model operating on a social network. However, since the agent-based model derived here is stochastic, it would be prohibitively expensive to solve a nonlinear optimization problem for new parameters based on mean trajectories obtained via an average of ABM simulations over the space of all feasible parameter values. Instead, we began by fitting the Phillips \textit{et al.} ODE model to the mean realization of the ABM generated with fixed network parameters and model parameters from the optimized Phillips \textit{et al.} model. Our motivation was to observe how the ODE model parameters would need to change in order to mimic an imposed network structure. Working backwards, we might then narrow our search for which of the ABM parameters would need to be adjusted to reproduce the ODE results.

Parameter estimation on the ODE model was conducted using the sequential least-squares quadratic programming (SLSQP) method through the Python library Scipy \cite{Virtanen2020}. The objective function $F$ for this optimization was the mean squared error between the ODE trajectory \[\mathbf{O}(t) = [S_{ode}(t), P_{ode}(t), A_{ode}(t), H_{ode}(t), R_{ode}(t)]\] and the pointwise mean ABM trajectory \[\mathbf{M}(t) = [S_{abm}(t), P_{abm}(t), A_{abm}(t), H_{abm}(t), R_{abm}(t)]\] of all compartments. To account for all compartments of the model, we constructed vectors at each discretized time $t_i$, resulting in a matrix of size $(1000, 5)$ for 1000 time-steps and 5 compartments. $F$ is then defined by

\begin{equation}
\label{eq:opt_loss}
F(\mathbf{M}, \mathbf{O}) = \frac{1}{1000}\sum\limits_{i=1}^{1000}\frac{1}{5}\sum\limits_{j=1}^{5}(M_{ij} - O_{ij})^2
\end{equation}

This optimization was then repeated for different types and parameterizations of network algorithms, each time searching for the optimal parameterization of the ODE model result $\mathbf{O}^*(t)$ with the smallest value of $F(\mathbf{M}, \mathbf{O}^*)$.

All optimizations started with initial conditions described in Tbl. \ref{tab:controls} located in \ref{appendix:base_parameters}. For all experiments, the optimization was run with a maximum 500 iterations and a target tolerance of $1*10^{-20}$ before the optimization algorithm stopped. Each reference ABM model was run 301 times with base parameters described in Tbl. \ref{tab:ba_opt_table} and network parameters described in \ref{sec:opt_controls} with the output trajectories averaged.

After the ODE model was fit to the mean ABM trajectory, we compared the original parameter values to the newly fitted values and used an inversion formula to approximate new parameter values that might be necessary to make the ABM match the original, data-driven Phillips \textit{et al.} ODE model trajectory. This works as follows: For a given parameter (e.g., $\alpha$) in the ABM and ODE models, we first find the value $k$ such that $\alpha_f = k\alpha_o$ where $\alpha_o$ is the original Philips \textit{et al.} value for $\alpha$ and $\alpha_f$ is the value of $\alpha$ when the ODE model is fit to the mean of a given ABM model. Inverting this function for parameter shift, we then have an approximation for the needed shift in the ABM parameter value to match the original ODE result, $\alpha^{abm} = \alpha_o/k=\alpha_o^2/\alpha_f$. For our discussion, we will refer to parameters such as $\alpha^{abm}$ as ``inverted'' parameters. To improve convergence of the optimization algorithm, we also imposed bounds on the fit values of the ODE model. Following the example of $\alpha$, we restricted the fit bound of $\alpha_f$ to $[\alpha_o*10^{-3}, \alpha_o]$. We use $\alpha_o$ as an upper-bound for $\alpha_f$ based on the intuition that the well-mixed ODE model should only decrease its value of network-dependent parameters in order to fit the sparser network structure used in the ABM.

Using these inverted parameters, we can compare visually how well the ABM might fit the ODE model results. This procedure was repeated for a variety of network topologies on the ABM. Statistical analysis was also conducted to compare how these inverted parameters varied according to network statistics relevant to the network generation algorithms we chose. These results will be described in Section \ref{sec:opt}.

\subsection{Experimental details}
In all cases, our ABM was run with a time step of $\Delta t = 0.01$ where $t$ is in years. In order to keep the model trajectory at 10 years (as was done in Phillips \textit{et al.} \cite{Phillips2021}), each model was run for 1000 time steps. After each time step, the population of individuals belonging to each class is recorded. 

In the case of all but the fully connected network model, we also record two primary network statistics: the average path length (APL) and the clustering coefficient (CC). These were used for network comparison. Other network statistics were recorded as well, including the mean and variance of node degree from each model. Each network statistic was recorded at the beginning of the simulation, before any alteration by the ABM. We found this recording strategy to be adequate after analysis of beginning and ending network statistics showed very little perturbation in values; we hypothesize that this is due to low death rates causing low reintroduction rates for each model (see base parameters in Tbl. \ref{tab:controls}) as well as our chosen methods of node reintroduction. 

Since our results rely on a mean trajectory of several ABM runs for a given network parameterization, we used a simple arithmetic mean to aggregate network statistics across multiple model realizations. These could then be compared between network generation algorithm parameterizations. These metrics were calculated using the NetworkX library in Python \cite{networkx}. The model was originally constructed in the NetLogo programming language \cite{wilensky1999netlogo}, but was later transferred to Python and implemented using the NetworkX library \cite{networkx}. All codes utilized in this research are publicly available from GitHub. The Phillips \textit{et al.} model is available at \url{https://github.com/mountaindust/Heroin_model} and all ABM-related codes used in this research are available with documentation at \url{https://github.com/owencqueen/SPAHR_Model}.

\section{Results}
\label{sec:Results}

\begin{figure}[h!]
    \centering
    \includegraphics[width=\linewidth]{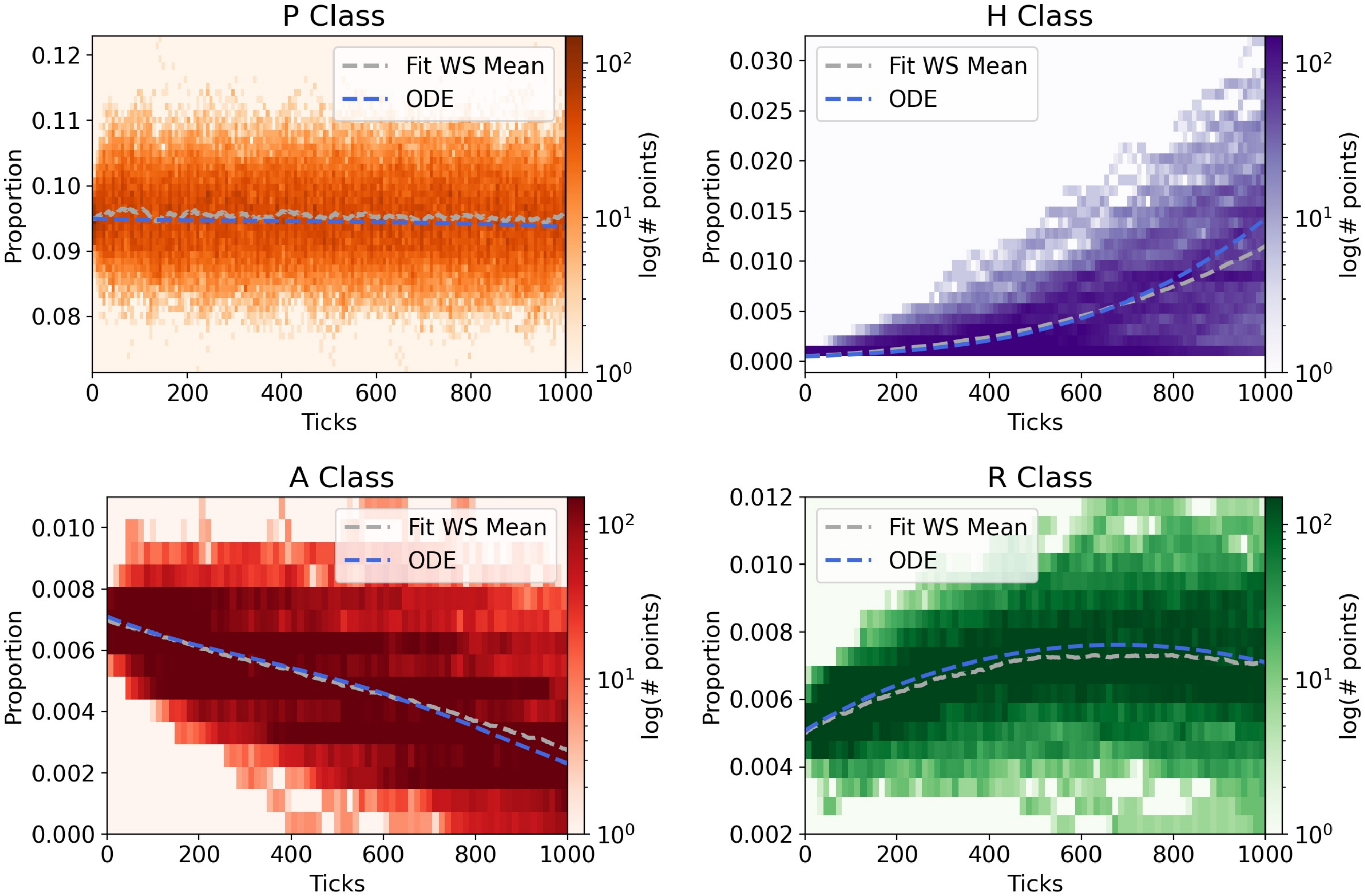}
    \caption{An example of boostrapping the ABM time-series distribution compared to the ODE model. The ABM distribution is presented by a heat map, on which we can take the point-wise mean value. 1 model tick = 0.01 years. This point-wise mean can be compared to the ODE model's trajectory for each compartment in the model. In this example, we use a WS model with $n = 2$ and $p = 0.2$. This results in trajectories that deviate significantly from the ODE model, as particularly evidenced in the $H$ class. This effect is due to the sparse network structure of the WS model used for the ABM in this example.}
    \label{fig:bootstrap}
\end{figure}

\subsection{Parameter estimation}
\label{sec:opt}

To analyze the effect of network structure and ABM stochasticity on the Phillips \textit{et al.} ODE model, it is necessary to fit one of these models to the results of the other. Since agent-based model results are the realization of a stochastic process while the Phillips \textit{et al.} ODE system is meant to represent a mean-field model, it is far easier to conduct parameter estimation on the ODEs, thereby fitting the Phillips \textit{et al.} to the ABM for a given social network structure. We then varied the network structure and examined how each parameter in the ODEs had to change in order to approximate the effect of the network on substance use dynamics. Details of the parameter estimation are described in Section \ref{sec:opt_methods}.

\subsubsection{Necessary network dynamics based on Phillips et al. results}

Initial parameter estimation focused on the ER models described in \ref{sec:opt_controls}. 

First, an optimization was attempted to find parameters that minimized Equation \ref{eq:opt_loss} for each tested network parameter set (\ref{sec:opt_controls}) of the Erd\H{o}s-R\'{e}nyi agent-based model, using a procedure as described in Section \ref{sec:opt_methods}. During this initial attempt, all network-dependent parameters (Tbl. \ref{tab:network_rates}) were optimized in the ODE model, with all other parameters constant.


When attempting to optimize all network-dependent parameters, the optimization could not produce a solution that reasonably fit the ABM to the ODE model. An \textit{ad hoc} ablation study was conducted to determine the source of improper fitting in the optimization of ODE parameters. Network dependence was removed for previously network-dependent parameters, and resulting fits were compared visually and quantitatively, by measuring the final loss value (Equation \ref{eq:opt_loss}) after convergence of the optimization algorithm. To convert parameters from network dependence to independence, we reverted to the Phillips \textit{et al.} definition of the parameters, as described in Tbl. \ref{tab:network_rates}.



\begin{table}[ht]
    \centering
    \begin{tabular}{ |p{2.5cm}|p{2.5cm}|p{2.5cm}| }
    \hline
         & $\theta_2$, $\theta_3$ vary, $H$ network & $\theta_2$, $\theta_3$ fixed, $H$ global \\
         \hline
         \hline
         MSE & $4.24 * 10 ^{-7}$ & $7.61 * 10^{-7}$\\
         $\theta_{1}^{abm}$ & 0.2294 & 0.6875 \\
         $\theta_{2}^{abm}$ & 236 & 0.236 (Fixed) \\
         $\theta_{3}^{abm}$ & 19700 & 19.7 (Fixed) \\
         $\beta_{P}^{abm}$ & $6.54 * 10^{-5}$ & $6.54 * 10^{-5}$ \\
         $\beta_{A}^{abm}$ & $8.78 * 10 ^ {-4}$ & $8.78 * 10 ^ {-4}$\\
         \hline
    \end{tabular}
    \caption{
    Fit values of parameters listed in Tbl. \ref{tab:network_rates} after an optimization procedure that fit the ODE model to an ABM model with a BA network ($m$ = 3). In the center column, results are shown for an optimization trial in which both $\theta_2$ and $\theta_3$ are allowed to vary and network-based definition of these rates is used (see Section \ref{sec:socialnetworks}). In the right-most column, we fix both $\theta_2$ and $\theta_3$ and use a network-independent, global definition of these parameters, i.e. the definition used in the Phillips model and mentioned in Tbl. \ref{tab:network_rates}. The top row lists the mean squared error (MSE) as defined in Equation \ref{eq:opt_loss} while each of the other rows list the inverted parameter value intended for use in the ABM.
    }
    \label{tab:theta23_fitting}
\end{table}

Upon conducting optimization experiments with the original network-dependent parameters, we observed that $\theta_2$ and $\theta_3$ values were decreasing drastically for the fit ODE, indicating that the inverted ABM parameters would be very high. This occurred especially in the BA model, where $\theta_{2}^{abm}$ and $\theta_{3}^{abm}$ values would often reach their upper-bounded values. A typical optimization trial with ABM definitions of $\theta_2$ and $\theta_3$ is shown in Tbl. \ref{tab:theta23_fitting}, where $\theta_{2}^{abm}$ and $\theta_{3}^{abm}$ become what we deemed as unreasonably large with respect to the model. To counter this effect, we removed network dependence for both $\theta_{2}^{abm}$ and $\theta_{3}^{abm}$ parameters, reverting back to the Phillips \textit{et al.} definitions as is shown in Tbl. \ref{tab:network_rates}. By reverting to the ODE definitions, we observed that our final loss value would increase by up to two-fold, but parameter values were much more controlled, with $\theta_1$ being the primary variable that controlled the goodness-of-fit. Tbl. \ref{tab:theta23_fitting} shows one example of this phenomena observed when removing network dependence from $\theta_2$ and $\theta_3$; similar patterns were observed across WS, BA, and ER models for a variety of model construction parameters. Therefore, the decision was made to only allow the parameters $\theta_1$, $\beta_A$, and $\beta_P$ to retain network dependence and vary during the optimization procedure.



We interpret this result as indicative of a lack of social network influence (in the sense of presence or absence of $H$ individuals in direct social contact) on initiation of heroin use for the $P$ and $A$ classes. Put another way, for individuals who may be developing an opioid use disorder based on prescription use ($P$ class) or already have an opioid use disorder ($A$ class), the social network simply has no bearing on how likely they are to move to the $H$ class compared to the relative prevalence of $H$ in the general population. Agents in these classes will develop heroin use disorder by seeking out their own, new access to heroin without the necessity of $H$ contacts within their existing network.

However, the case of $S$ is different: Individuals who are not actively using or recovering from opioids develop opioid or heroin use disorder through direct, social connections to $P$, $A$, or $H$ agents. In terms of the model parameters, this means that $\beta_P$, $\beta_A$, and $\theta_1$ remain network-dependent pathways in the ABM.

\subsection{Analysis of Network Statistics}
\label{results:analyze_net_stats}

\begin{table}[t]
    \centering
    \begin{tabular}{ |p{2cm} p{2cm} p{2.5cm} p{2.5cm}| }
    \hline
         x & y & $r$ & $p$-value \\
         \hline
         \hline
         APL & $A_f$ & 0.158008 & 5.79e-30\\
         APL & $H_f$ & -0.321327 & 3.13e-123 \\
         APL & $A_f+H_f$ & -0.331160 & 3.22e-131 \\
         \hline
         CC & $A_f$ & -0.091338 & 5.92e-11 \\
         CC & $H_f$ & 0.220816 & 1.51e-57 \\
         CC & $A_f+H_f$ & 0.233770 & 1.81e-64 \\
         \hline
         DMean & $A_f$ & -0.088814 & 1.96e-10 \\
         DMean & $H_f$ & 0.217396 & 8.60e-56 \\
         DMean & $A_f+H_f$ & 0.230548 & 1.05e-62 \\
         \hline
         DVar & $A_f$ & -0.092097 & 4.10e-11 \\
         DVar & $H_f$ & 0.223598 & 5.36e-59 \\
         DVar & $A_f+H_f$ & 0.236856 & 3.50e-66 \\
         \hline
    \end{tabular}
    \caption{Pearson’s correlation coefficient ($r$) between a variety of statistics for each network and modeling outcome for model simulations on an Erd\H{o}s-R\'{e}nyi random network (reference \ref{sec:opt_controls} for parameters of models tested). APL is average path length, CC is clustering coefficient, DMean is mean degree of all nodes in the network, DVar is variance of degree between all nodes in the network, and $A_f$ and $H_f$ is the final proportion of $A$ and $H$ individuals respectively. The $p$-value is also given for the computation of $r$ in each relationship.}
    \label{tab:net_stats}
\end{table}

Various experiments were performed in order to analyze network statistics against baseline modeling outcomes derived from the Phillips \textit{et al.} ODE model. Certain variables were kept constant throughout; these control values, based on values obtained by Phillips \textit{et al.}, are shown in Tbl. \ref{tab:controls}.

To begin, various network statistics were tested against a variety of Erd\H{o}s-R\'{e}nyi (ER), Barab\'{a}si-Albert (BA), and Watts-Strogatz (WS) models for their relative importance to the final number of $A$ and $H$ individuals, denoted $A_f$ and $H_f$ respectively. The parameter values for each model considered in this analysis are detailed in \ref{sec:opt_controls}. We repeated each simulation 300 times for each chosen parameter value, with each simulation starting from the initial conditions $S_0$, $P_0$, $A_0$, $H_0$, and $R_0$ as shown in Tbl. \ref{tab:controls} in the Appendix and then run for 1000 time steps. We found that the change in network statistics from beginning to end of each simulation run was insignificant (our reintroduction algorithms worked as intended in this regard); therefore, statistics for the network topology were recorded at the beginning of each simulation.

\begin{figure}[ht]
    \centering
    \includegraphics[width=\linewidth]{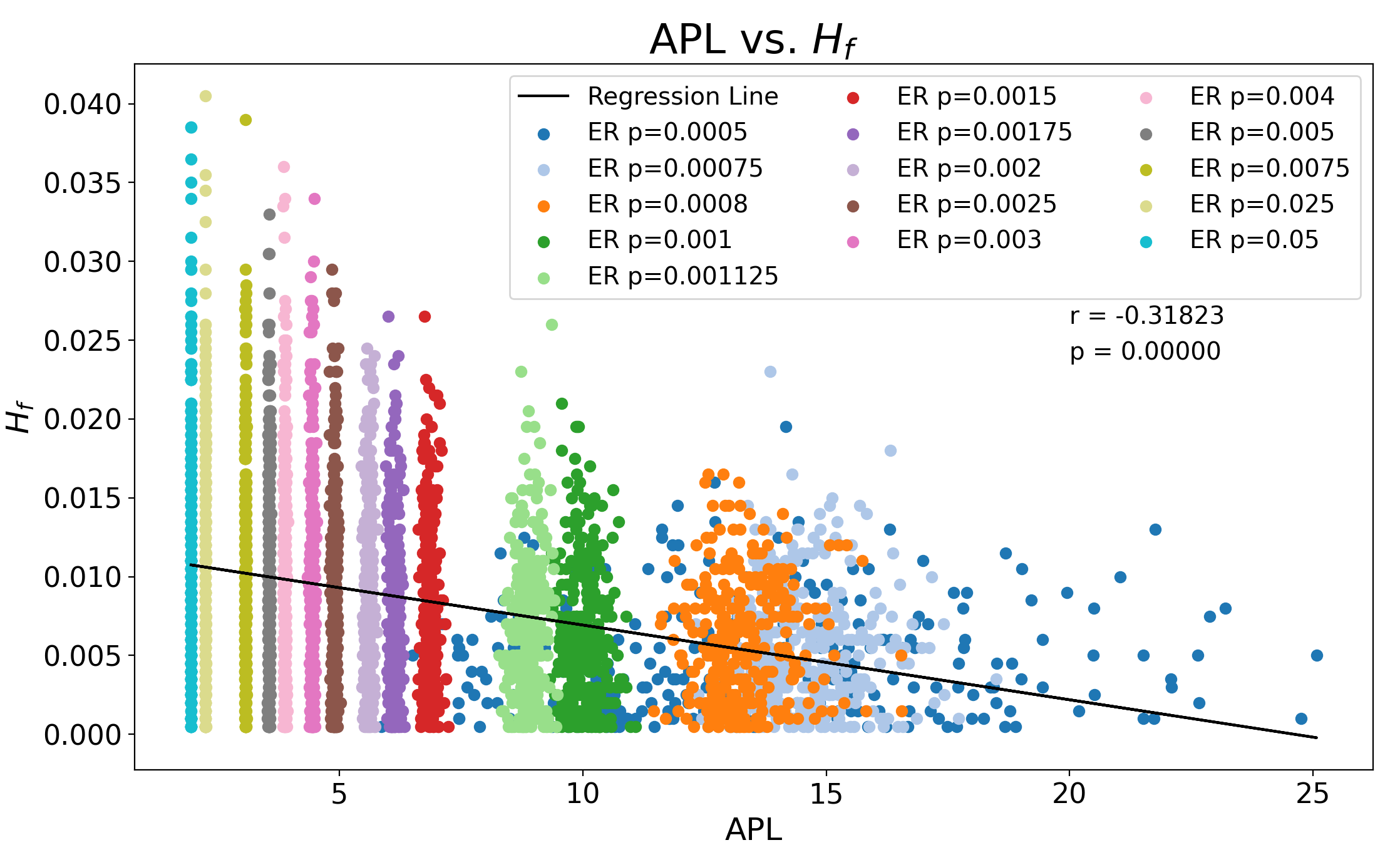}
    \caption{ $H_f$ plotted against beginning APL value for each of several runs of ER models with varying $p$ parameters. A least-squares regression line is also shown to emphasize the negative correlation between APL and $H_f$. Each model parameter was chosen to display a wide range of APL values. In the text below the legend, the Pearson's correlation coefficient ($r$) and the p-value for this $r$ statistic is displayed. The p-value is very low (truncated to 0), meaning that this is a statistically significant correlation coefficient.}
    \label{fig:aplhf_ER}
\end{figure}

The resulting correlations for the ER network statistics versus final values of $A$ and $H$ are shown in Tbl. \ref{tab:net_stats}. Note that the network statistics are not independent of each other, so their correlations should be compared for relative importance rather than taken in isolation. One can see that on this relative basis, average path length (APL) appears to be very important for the ER models in strength of correlation with $H_f$ and $A_f+H_f$. This relationship is negative, meaning that the more sparse the network becomes (higher APL), the more $H_f$ is expected to decrease. Note that none of the statistics tested showed a strong correlation with $A_f$. We have visualized the data and regression line relating APL and $H_f$ in Fig. \ref{fig:aplhf_ER}.

Similar correlation analyses for BA and WS models are shown in Tbls. \ref{tab:BA_net_stats} and \ref{tab:WS_net_stats} in the appendix. Across all models, APL seems to produce a consistently strong correlation against $H_f$ values, but for BA and WS, this correlation is weaker than for ER models. We hypothesize that this is because the exact location of $A$ and $H$ nodes within the non-trivial network structure matters far more than in an ER network, where the network structure can be thought of as more homogeneous. For WS, the degree mean and degree variance show the strongest correlation to $H_f$ values by far. However, this can be explained by the nature of the WS generation algorithm. The WS algorithm starts with all nodes having identical degrees, and only by rewiring, which is more frequent with a higher value for the $p$ parameter, would this degree change. Therefore, a higher $p$ parameter would cause more variance in the degree and also tends to decrease APL and CC (clustering coefficient) \cite{WS_network}. Likewise, the mean degree over all nodes in the network is directly correlated with the $n$ parameter, and this parameter also has a direct effect on the value of the APL and CC for that network \cite{WS_network}.

\subsubsection{Comparing Models with Similar APLs}

In order to understand the significance of each of these statistics for predicting $H_f$, an analysis was conducted using two network models with approximately equal APL. The goal was to hold APL constant and evaluate how changes in CC affect the modeling outcomes. APL was chosen as the fixed parameter for this analysis due to its strength in correlation for $H_f$, as previously discussed.

The first models tested were Erd\H{o}s-R\'{e}nyi with $p=0.0026$ and Watts-Strogatz with $n = 8$, $p=0.2$. Model parameters were chosen in order to produce networks with APLs with a difference of less than 0.001. In addition, these Erd\H{o}s-R\'{e}nyi and Watts-Strogatz network structures were chosen due to strong differences in their structure, evidenced by the difference in mean and variance of the degrees of nodes within each of these networks as shown in Tbl. \ref{tab:ks_es_erws}. Each model was run 2000 times as previously described. After each run, the $H_f$ and APL values were recorded.  

\begin{figure}[h]
    \centering
    \includegraphics[width=\linewidth]{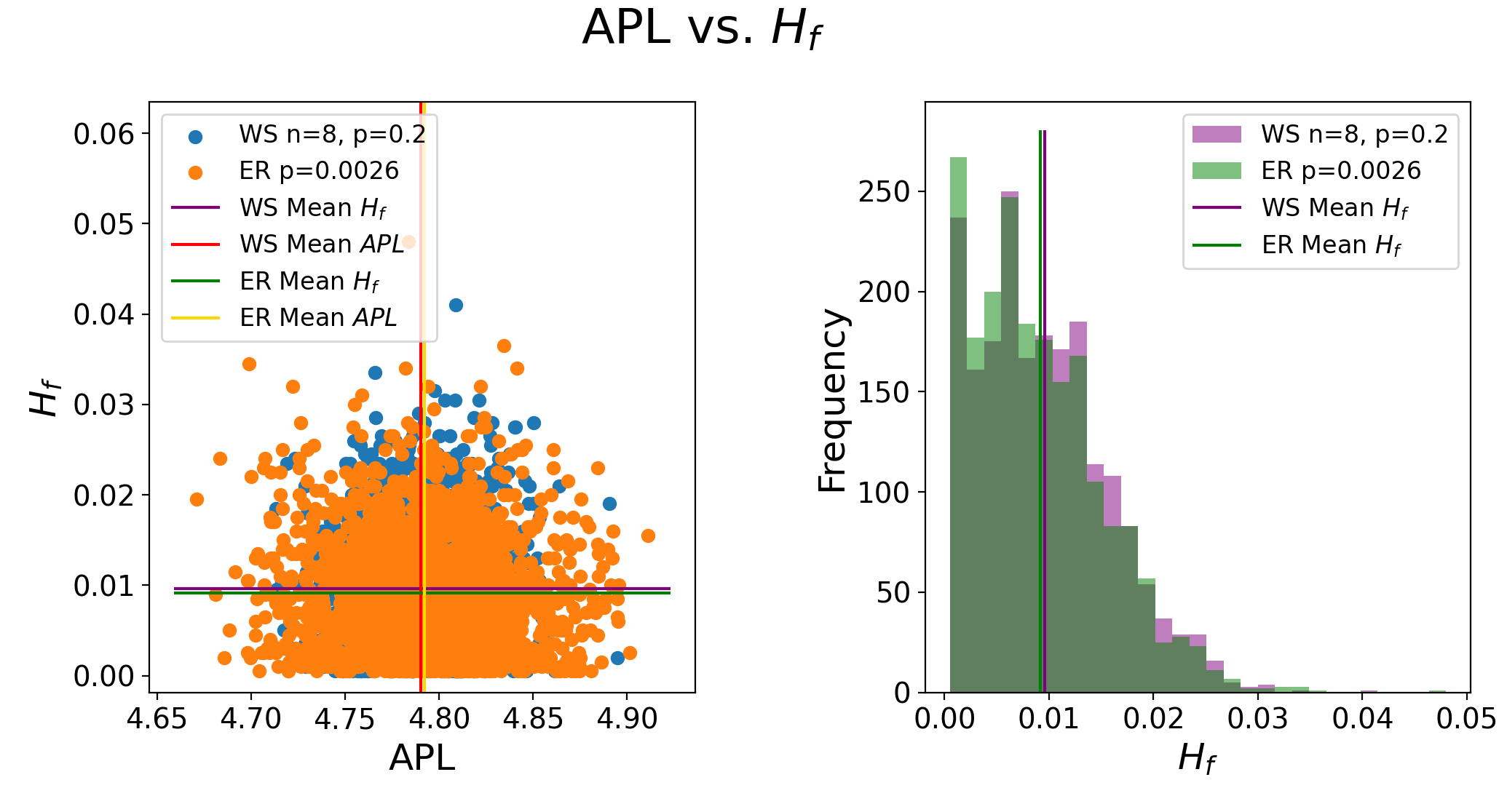}
    \caption{ Comparing the Erd\H{o}s-R\'{e}nyi model (p = 0.0026) and Watts-Strogatz (n = 8, p = 0.2) model on the agent-based model. On the left, average path length (APL) is plotted against the final proportion of $H$ individuals ($H_f$) for each run. On the right, a plot is shown of the distribution of $H_f$ for each of the model runs. The mean APL of the ER model was 4.44412 and the mean APL of the WS model is 4.44062, yielding a difference of 0.00350.}
    \label{fig:ws_er_comp_APL}
\end{figure}

Two statistical tests were performed to quantify similarity between distributions of $H_f$ from these models: the two-sample Kolmogrov-Smirnov (KS) test and the two-sample Epps-Singleton (ES) test, which is the discrete analog for KS test. The null hypothesis $H_0$ of each of these tests is that both samples have equivalent underlying distributions. Model results are visualized in Fig. \ref{fig:ws_er_comp_APL} with the corresponding statistics given in Tbl. \ref{tab:ks_es_erws}. APL and CC are both continuous values, thus they should be evaluated using the KS test. However, $H_f$ is a discrete distribution for the fixed population size within the model, so it should be evaluated using the ES test. For completeness, both KS and ES tests were run on each of these statistics.

In the left plot of Fig. \ref{fig:ws_er_comp_APL}, APL distributions seem very similar, but the KS test strongly suggests different APL distributions between the ER and WS models (Tbl. \ref{tab:ks_es_erws}). There is a weak statistical signal indicating possible correspondence between the $H_f$ distribution produced by the various runs of these two models. As expected, there is strong statistical evidence against similarity in clustering coefficient distributions across these two models - a result of the differences in generation of the networks underlying these models.

\begin{table}
    \centering
    {\footnotesize
    \begin{tabular}{|p{1.2cm}||p{1.5cm}|p{1.5cm}||p{1.3cm}|p{1.7cm}||p{1.6cm}|p{1.7cm}|}
        \hline
         & \multicolumn{2}{|c||}{\textbf{Mean Values}} & \multicolumn{2}{|c||}{\textbf{KS}} & \multicolumn{2}{|c||}{\textbf{ES}} \\
         \hline
          Statistic & ER & WS & Value & $p$-value & Value & $p$-value\\
         \hline
         APL & 4.7921 & 4.7903 & 0.11454 & 6.8773e-12 & 310.94 & 4.7329e-66 \\
         CC & 0.0025180 & 0.34317 & 1.0 & 0.0 & 4.6987e11 & 0.0 \\
         \hline
         DMean & 5.2259 & 8.0008 & 1.0 & 0.0 & 2.1651e10 & 0.0 \\
         DVar & 5.0533 & 2.3704 & 1.0 & 0.0 & 9.8268e6 & 0.0 \\
         \hline
         $H_f$ & 0.0091352 & 0.0096001 & 0.043825 & 0.042269 & 9.1538 & 0.057367 \\
         \hline
    \end{tabular}
    \caption{Two-sample Kolmogrov-Smirnov (KS) and Epps-Singleton (ES) tests of two samples, one from ER ($p=0.0026$) and one from WS ($n=8$, $p=0.2$). }
    \label{tab:ks_es_erws}\
    }
\end{table}

\subsection{Determining how average-path length relates to model parameters}

\begin{table}[t]
    \centering
    \begin{tabular}{ |p{1.6cm} p{1.5cm} p{1.7cm} p{1.5cm} p{1.5cm} p{2cm}| }
    \hline
    $p$ & APL & $\beta_{A}^{abm}$ & $\beta_{P}^{abm}$ & $\theta_{1}^{abm}$ & final loss \\
    \hline
    0.00075 & 14.642 & 0.000878 & 6.54e-05 & 0.47798 & 5.24e-07 \\ 
    0.0008 & 13.28 & 0.000878 & 6.54e-05 & 0.43961 & 3.16e-07 \\ 
    0.001 & 10.018 & 0.000878 & 6.54e-05 & 0.41369 & 3.72e-07 \\ 
    0.001125 & 8.8839 & 0.000878 & 6.54e-05 & 0.34503 & 5.91e-07 \\
    \hline
    0.0015 & 6.846 & 0.000878 & 6.54e-05 & 0.31686 & 4.65e-07 \\ 
    0.00175 & 6.1288 & 0.000878 & 6.54e-05 & 0.30546 & 3.62e-07 \\ 
    0.002 & 5.6099 & 0.000878 & 6.54e-05 & 0.29302 & 4.11e-07 \\ 
    0.0025 & 4.8977 & 0.000878 & 6.54e-05 & 0.28086 & 4.08e-07 \\
    \hline
    0.003 & 4.4434 & 0.000878 & 6.54e-05 & 0.28186 & 2.95e-07 \\ 
    0.004 & 3.8839 & 0.000878 & 6.54e-05 & 0.25782 & 4.81e-07 \\ 
    0.005 & 3.5613 & 0.000878 & 6.54e-05 & 0.25263 & 3.54e-07 \\ 
    0.0075 & 3.0773 & 0.000878 & 6.54e-05 & 0.24445 & 5.02e-07 \\
    \hline
    0.01 & 2.8306 & 0.000878 & 6.54e-05 & 0.24367 & 4.29e-07 \\ 
    0.025 & 2.2544 & 0.000878 & 6.54e-05 & 0.24158 & 6.2e-07 \\ 
    0.03 & 2.1304 & 0.000878 & 6.54e-05 & 0.24157 & 3.73e-07 \\ 
    0.05 & 1.9564 & 0.000878 & 6.54e-05 & 0.245 & 3.41e-07 \\
    \hline
    \end{tabular}
    \caption{Optimization results for a variety of ER models. $\theta_{1}^{abm}$ is the ABM, ''inverted'' value as previously described in Section \ref{sec:opt_methods}. The final loss is the value of mean squared error between the ABM projection mean and the ODE model projection for every class. These relationships are visualized in Fig. \ref{fig:apl_v_theta}.}
    \label{tab:er_opt_table}
\end{table}

Using a parameter sweep, we analyzed the affect of average-path length (APL) on the outcome of the parameter estimation procedure described in Section \ref{sec:opt_methods}. APL is the focus of these analysis due to the strong correlations discovered between APL and model outcome in the analyses described in Section \ref{results:analyze_net_stats}. 


We found that $\theta_1$ is the only parameter which exhibits any significant change with respect to APL. Tbls. \ref{tab:er_opt_table}, \ref{tab:ba_opt_table}, and \ref{tab:ws_opt_table} show optimization results for changes in parameters for a variety of ER, BA, and WS models, respectfully (BA and WS tables are located in the appendix).

\subsubsection{APL as a Predictor of $\theta_1$}

\begin{figure}[h]
    \centering
    \includegraphics[width=\linewidth]{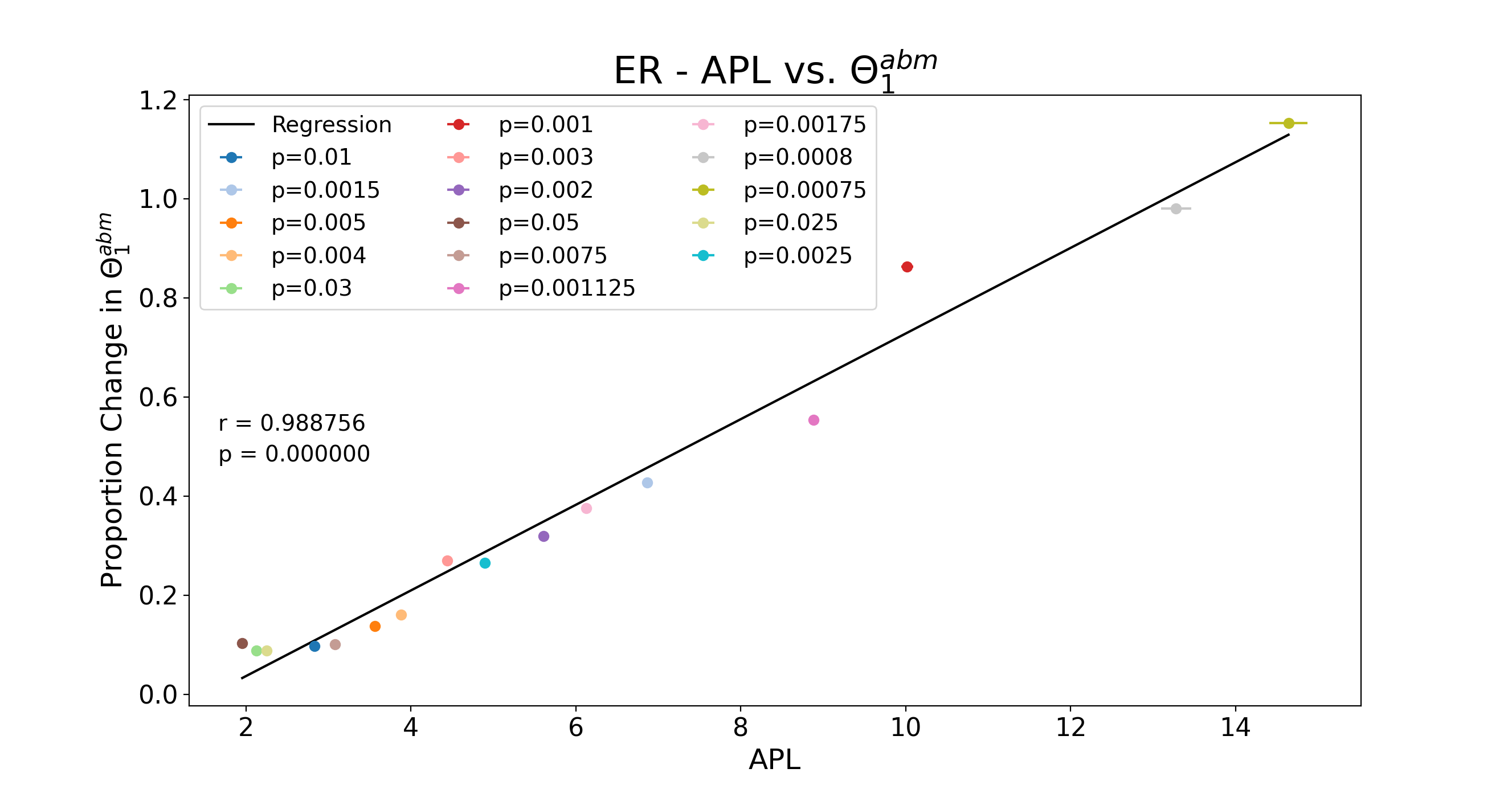}
    \caption{ APL vs. inverted $\theta_1$ values for optimizations ran on ER models with varying $p$ values. Each point represents the derived $\theta_1$ value plotted against the average APL over 300 model runs, with each network in these 300 runs being generated with the same $p$ parameter. Horizontal bars are shown around each point corresponding to a 95\% confidence interval for that APL value. The Pearson's correlation coefficient ($r$) and the $p$-value for that $r$ calculation is also displayed. Separate analyses are conducted with WS (Fig. \ref{fig:ws_apl_v_theta}) and BA (Fig. \ref{fig:ba_apl_v_theta}) models.}
    \label{fig:apl_v_theta}
\end{figure}

APL and the resulting change in $\theta_{1}^{abm}$ (from an original value of 0.222) appear to have a strong linear relationship (Pearson r of 0.988756, p $< 10^{-6}$) for ER network structures. This relationship is plotted in Fig. \ref{fig:apl_v_theta}. The same analysis is shown for BA models in Fig. \ref{fig:ba_apl_v_theta} and for WS models in Fig. \ref{fig:ws_apl_v_theta}. The Pearson's correlation coefficient for the BA network models was found to be 0.719206 with a p-value of 0.008382, and for the 
WS network models, it was 0.845941 with a p-value of ($< 10^{-6}$). 

\section{Conclusions}
Our study utilized an assumption that the rates of change given in the Phillips \textit{et al.} \cite{Phillips2021} model represent the mean of an underlying Poisson process. This assumption was used to convert the ordinary differential equations into a stochastic, individual-based model which could then be combined with a social network. Social networks were stochastically generated from three different algorithms and were static for each simulation except for when a node underwent a death process and was subsequently added back into the network using a method similar to the original generation algorithm. The result is a random process which can be used to bootstrap a time series distribution (see Fig. \ref{fig:bootstrap}) and explore the effects of different social network dynamics and structures.

Since relaxation of the well-mixed assumption of the Phillips \textit{et al.} equations could be expected to decrease the transmissibility of opioid use disorder, and the Phillips \textit{et al.} time-series results were fit to data from Tennessee, an obvious question was to explore how the rate parameters in the agent-based model would need to change to recover the time-series results of Phillips \textit{et al.} in the presence of a non-trivial social network structure. The computational cost of fitting parameters based on a bootstrapped mean of many stochastic realizations of the ABM to the Phillips \textit{et al.} time series result is prohibitive, so instead parameters were estimated by fitting the Phillips \textit{et al.} model to a single bootstrapped mean of the ABM and then noting the change in parameters between the ODE model based on TN data and the ABM fitted ODE model. This change was reversed to estimate an ABM model that would approximate the Phillips \textit{et al.} result based on TN data.

We quickly discovered that our original assumption that the $P\rightarrow H$ and $A\rightarrow H$ transitions (with parameters $\theta_2$ and $\theta_3$) depended on $H$ neighbors in the social network resulted in outlandish values for $\theta_2$ and $\theta_3$ when compared to data. Upon further reflection, it made sense that individuals who were already using prescription opioids in one way or another would likely not require existing social contacts using heroin or fentanyl in order to initiate heroin or fentanyl use. The model was changed so that these transitions were based on the total proportion of $H$ in the network rather than just neighbors, and $\theta_2$ and $\theta_3$ were fixed at their Phillips \textit{et al.} reported values. The resulting ABM model was capable of reproducing the Phillips \textit{et al.} result quite well. Additionally, we discovered that $\theta_1$ (representing the rate for the $S\rightarrow H$ transition) was the only parameter needing adjustment due to adding a social network.

Seeking to further explore the relationship between social network structure and $\theta_1$, we leveraged different network generation algorithms in order to vary a common collection of social network statistics, including average path length (APL), clustering coefficient (CC), degree mean (DMean), and degree variance (DVar). We found that a linear relationship with average path length was capable of explaining a large portion of the variance in $\theta_1$ due to the social network ($r=0.989$ for Erd\H{o}s-R\'{e}nyi, $r=0.947$ for Barab\'{a}si-Albert, and $r=0.846$ for Watts-Strogatz). We hope that this information may be used to infer the ``infectivity'' of heroin and fentanyl (specifically, heroin or fentanyl initiation by opioid naive individuals caused by social contact with heroin or fentanyl users) in communities where average path length of social contact can be estimated. Quantifying this information across different types of communities may shed significant light on factors that raise or lower risk for heroin and fentanyl use, thereby providing targets for management and further quantitative study.

\section*{CRediT author statement}
\textbf{Owen Queen:} Methodology, Software, Investigation, Formal analysis, Writing - Original draft, Visualization. \textbf{Vinny Jodoin:} Conceptualization, software. \textbf{Leigh Pearcy:} Software, Investigation, Writing - Review \& Editing. \textbf{Christopher Strickland:} Conceptualization, Methodology, Writing - Original Draft and Review \& Editing, Supervision, Project administration.

\section*{Acknowledgements}
The authors would like to acknowledge Suzanne Lenhart, Sheridan Payne, and Meagan Todd for their contributions to the 2019 NIMBioS-NSA REU project ``Modeling networking and the opioid epidemic'' which inspired the work in this manuscript.\\
Funding: dissemination of the results of this work was supported by the Simons Foundation Collaboration Grants for Mathematicians.

 \bibliographystyle{elsarticle-num} 
 \bibliography{references_all}






\appendix

\section{Model Parameter Values}
\label{appendix:base_parameters}

\begin{table}[h]
    \centering
    \begin{tabular}{|c c|c c|}
        \hline
        Statistic & Value & Statistic & Value\\
        \hline \hline
        $n$ & 2000 & $\theta_3$ & 19.7 \\
        Time Steps & 1000 & $\nu$ & 0.000531 \\
        $\Delta t$ & 0.01 & $\mu_H$ & 0.0466 \\
        $\mu$ & 0.0071 & $\sigma$ & 0.102 \\
        $\beta_A$ & 0.000878 & $\alpha$ & 0.27 \\
        $\beta_P$ & 0.0000654 & $\mu_A$ & 0.00883 \\
        $\theta_1$ & 0.222 & $S_0$ & 0.892365\\
        $\gamma$ & 0.00505 & $P_0$ & 0.095\\
        $\theta_2$ & 0.236 & $A_0$ & 0.0071\\
        $\epsilon$ & 2.53 & $H_0$ & 0.000465\\
        $\zeta$ & 0.198 & $R_0$ & 0.00507\\
        \hline
    \end{tabular}
    \caption{Control parameters used for every trial in the analysis of network statistics. Each parameter is as defined by Phillips et al \cite{Phillips2021}, with $n$ denoting the number of agents in the network. The number of agents in some class $X \in \{S, P, A, H, R\}$ is found by multiplying $nX(t)$ and rounding to the nearest integer. For example, the total number of $P$ individuals is $nP_0 = (2000)(0.095) = 190$.}
    \label{tab:controls}
\end{table}

Unless otherwise specified, each experiment in this study utilized the control parameters shown in Tbl. \ref{tab:controls}. 


\section{Correlation of Network Statistics with Model Outcomes}
\label{appendix:correlation}

Abbreviations are used to represent each statistic analyzed throughout this section. APL is average path length, CC is clustering coefficient, DMean is mean degree of all nodes in the network, DVar is variance of degree between all nodes in the network, and $A_f$ and $H_f$ is the final proportion of $A$ and $H$ individuals respectively. The $p$-value is also given for the computation of $r$ in each relationship. See Tbls. \ref{tab:BA_net_stats}-\ref{tab:WS_net_stats} for results.

\begin{table}
    \centering
    \begin{tabular}{ |p{2cm} p{2cm} p{2.5cm} p{2.5cm}| }
    \hline
         x & y & $r$ & $p$-value \\
         \hline
         \hline
         APL & $A_f$ & 0.0569716 & 6.13e-4\\
         APL & $H_f$ & -0.0937874 &  1.63e-8\\
         APL & $A_f+H_f$ & -0.0935757 &  1.76e-8\\
         \hline
         CC & $A_f$ & -0.0425750 & 0.0104965 \\
         CC & $H_f$ & 0.0699229 & 2.60e-5 \\
         CC & $A_f+H_f$ & 0.0697370 & 2.73e-5 \\
         \hline
         DMean & $A_f$ & -0.0368691 & 0.0267041 \\
         DMean & $H_f$ & 0.0653898 & 8.40e-5 \\
         DMean & $A_f+H_f$ & 0.0660425 & 7.12e-5 \\
         \hline
         DVar & $A_f$ & -0.0305392 & 6.65e-2 \\
         DVar & $H_f$ & 0.0545566 & 1.04e-3 \\
         DVar & $A_f+H_f$ & 0.0551633 & 9.11e-4 \\
         \hline
    \end{tabular}
    \caption{Pearson’s correlation coefficient ($r$) between a variety of statistics for each network and modeling outcome for model simulations on an Barab\'{a}si-Albert random network (reference \ref{sec:opt_controls} for parameters of models tested). }
    \label{tab:BA_net_stats}
\end{table}

\begin{table}
    \centering
    \begin{tabular}{ |p{2cm} p{2cm} p{2.5cm} p{2.5cm}| }
    \hline
         x & y & $r$ & $p$-value \\
         \hline
         \hline
         APL & $A_f$ & 0.0147997 & 3.20e-1\\
         APL & $H_f$ & -0.0841629 & 1.47e-8\\
         APL & $A_f+H_f$ & -0.0984920 & 3.25e-11\\
         \hline
         CC & $A_f$ & -0.0472457 & 1.49e-3 \\
         CC & $H_f$ & 0.0835066 & 1.90e-8 \\
         CC & $A_f+H_f$ & 0.0834411 & 1.94e-8  \\
         \hline
         DMean & $A_f$ & -0.0669150 & 6.75e-6 \\
         DMean & $H_f$ & 0.163304 & 2.24e-28 \\
         DMean & $A_f+H_f$ & 0.174352 & 3.61e-32 \\
         \hline
         DVar & $A_f$ & -0.0663410 & 8.09e-6 \\
         DVar & $H_f$ & 0.161539 & 8.58e-28 \\
         DVar & $A_f+H_f$ & 0.172401 & 1.76e-31 \\
         \hline
    \end{tabular}
    \caption{Pearson’s correlation coefficient ($r$) between a variety of statistics for each network and modeling outcome for model simulations on an Watts-Strogatz random network (reference \ref{sec:opt_controls} for parameters of models tested). }
    \label{tab:WS_net_stats}
\end{table}

\section{Comparing Networks with Similar APLs}
\label{appendix:apl_compare}

\begin{figure}[h]
    \centering
    \includegraphics[width=\linewidth]{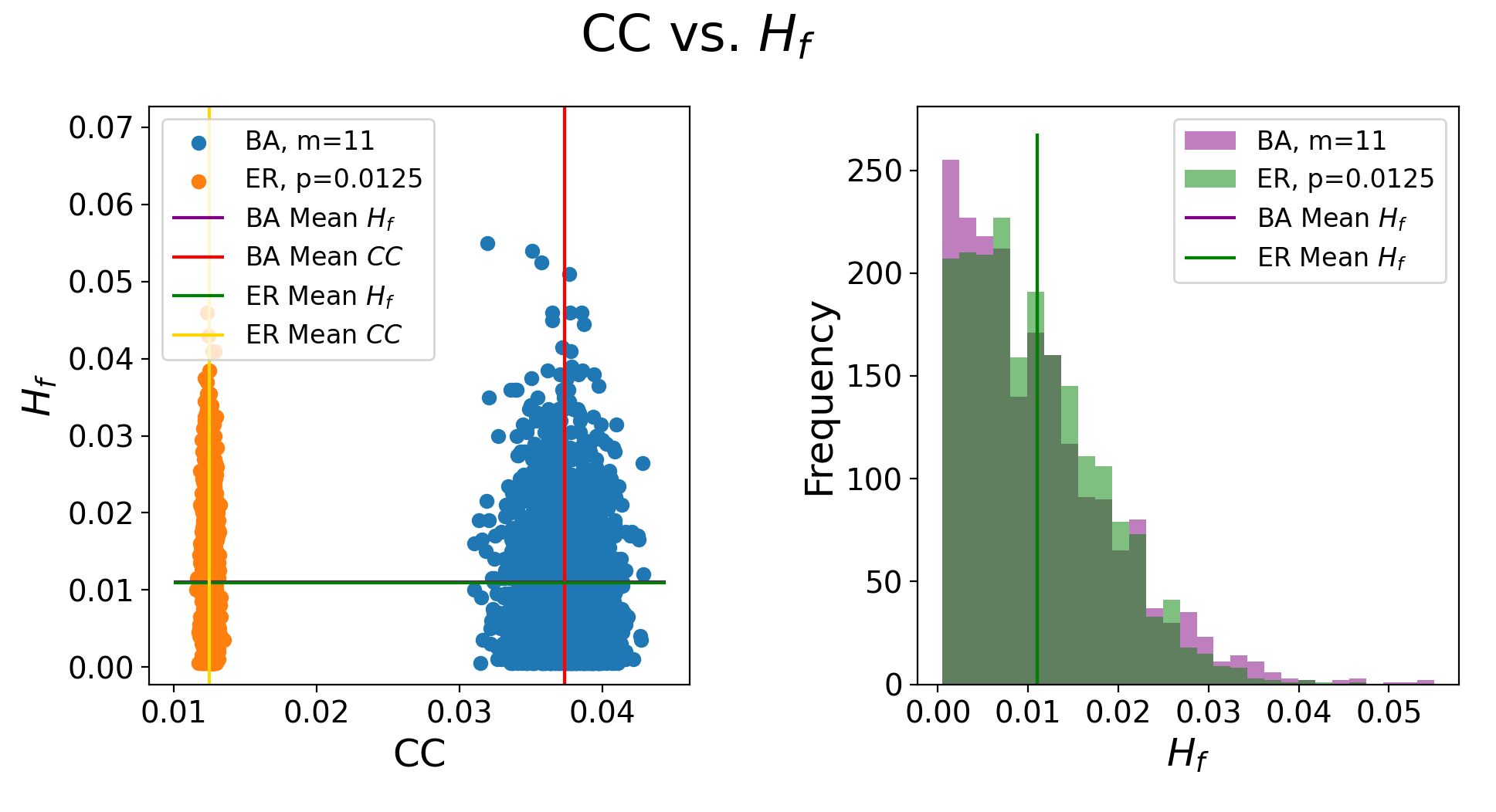}
    \vspace{-0.8cm}
    \caption{Comparing the effects of the Erd\H{o}s-R\'{e}nyi network (p = 0.0125) and Barab\'{a}si-Albert network (m = 11) on the ABM. On the left, clustering coefficient (CC) is plotted against the final proportion of $H$ individuals ($H_f$) for each realization. On the right, a plot is shown of the distribution of $H_f$ for each model realization. The two network models produce different CC values while retaining similar distributions for $H_f$, similar to Fig. \ref{fig:ws_er_comp_CC}.}
    \label{fig:er_ba_comp_CC}
    \includegraphics[width=\linewidth]{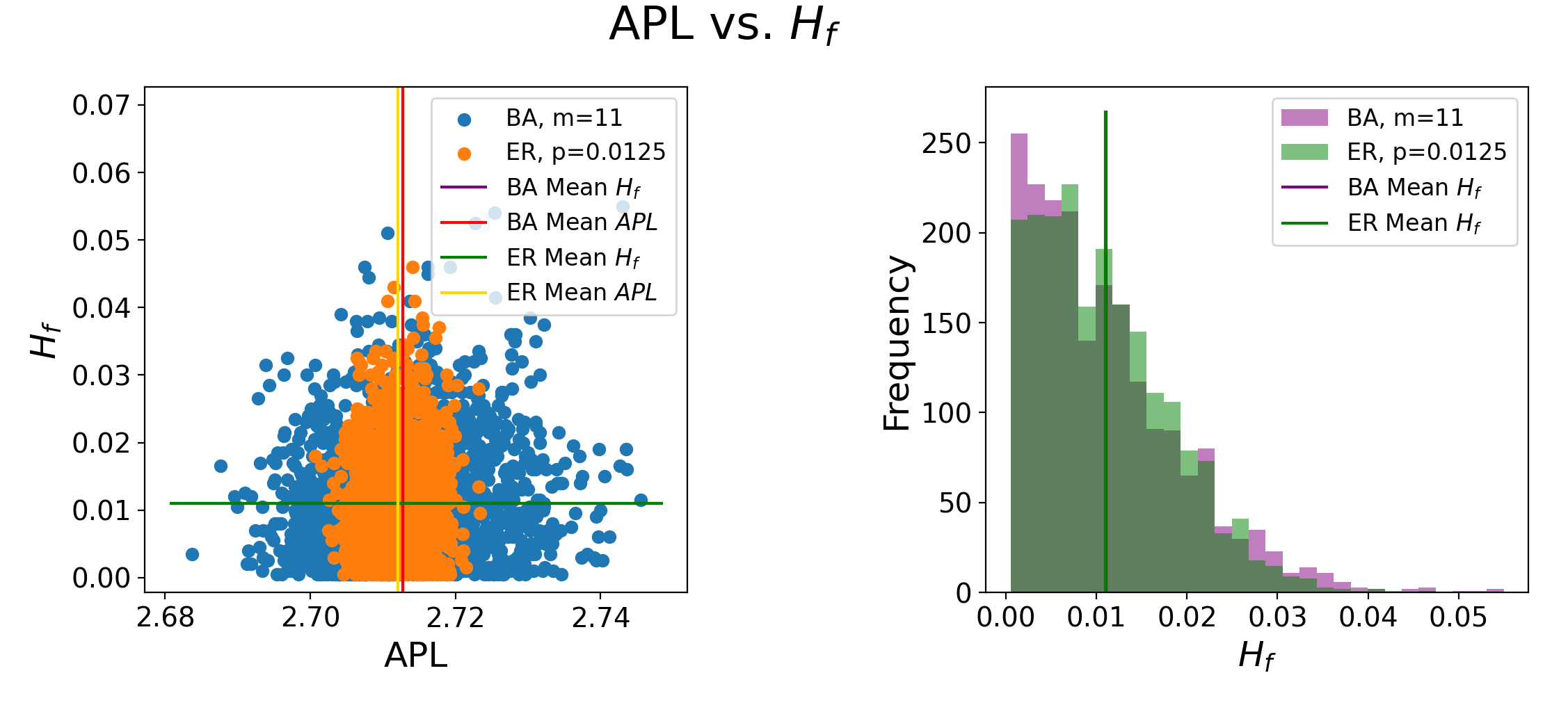}
    \vspace{-0.8cm}
    \caption{Comparing the effects of the Erd\H{o}s-R\'{e}nyi network (p = 0.0125) and Barab\'{a}si-Albert network (m = 11) on the ABM. On the left, average path length (APL) is plotted against the final proportion of $H$ individuals ($H_f$) for each realization. On the right, a plot is shown of the distribution of $H_f$ for each model realization. While the distributions of APL have different variance, and therefore different distributions (see Tbl. \ref{tab:ks_es_erba}), similar values of APL yield a similar distribution of values for $H_f$.} 
    \label{fig:er_ba_comp_APL}
\end{figure}

See results in Figs. \ref{fig:ws_er_comp_CC} - \ref{fig:er_ba_comp_CC} and Tbl. \ref{tab:ks_es_erba}.

\clearpage

\begin{minipage}{\linewidth}
    \centering
    \footnotesize
    \begin{tabular}{|p{1.3cm}||p{1.4cm}|p{1.5cm}||p{1.3cm}|p{1.7cm}||p{1.3cm}|p{1.8cm}|}
        \hline
         & \multicolumn{2}{|c||}{\textbf{Mean Values}} & \multicolumn{2}{|c||}{\textbf{KS}} & \multicolumn{2}{|c|}{\textbf{ES}} \\
         \hline
         Statistic & ER & BA & Test Statistic & $p$-value & Test Statistic & $p$-value\\
         \hline
         APL & 2.7121 & 2.7128 & 0.23330 & 1.5301e-48 & 1421.2 & 1.7134e-306 \\
         CC & 0.012501 & 0.037332 & 1.0 & 0.0 & 1.1144e8 & 0.0 \\
         \hline
         DMean & 24.988 & 20.398 & 1.0 & 0.0 & 2.4957e7 & 0.0 \\
         DVar & 24.640 & 390.75 & 1.0 & 0.0 & 3.9330e8 & 0.0 \\
         \hline
         $H_f$ & 0.010986 & 0.0018989 & 0.039595 & 0.0824647 & 31.145 & 2.8592e-6 \\
         \hline
    \end{tabular}
    \captionof{table}{Two-sample Kolmogrov-Smirnov (KS) and Epps-Singleton (ES) tests of 2000 realizations from two models, one using an ER network ($p=0.0125$) and the other BA ($m=11$). Lower p-values indicate statistical significance suggesting that the two sample sets were drawn from different underlying distributions. The tested network statistics from each model are shown in the leftmost column. The results highly suggest different distributions for APL, CC, and $H_f$ between the two networks. However, we note that the mean $H_f$ value was almost equivalent between the two models even while their distributions are different (see Fig. \ref{fig:er_ba_comp_APL}).}
    \label{tab:ks_es_erba}
    
    \vspace{1cm}
    
    \normalsize
    \includegraphics[width=\linewidth]{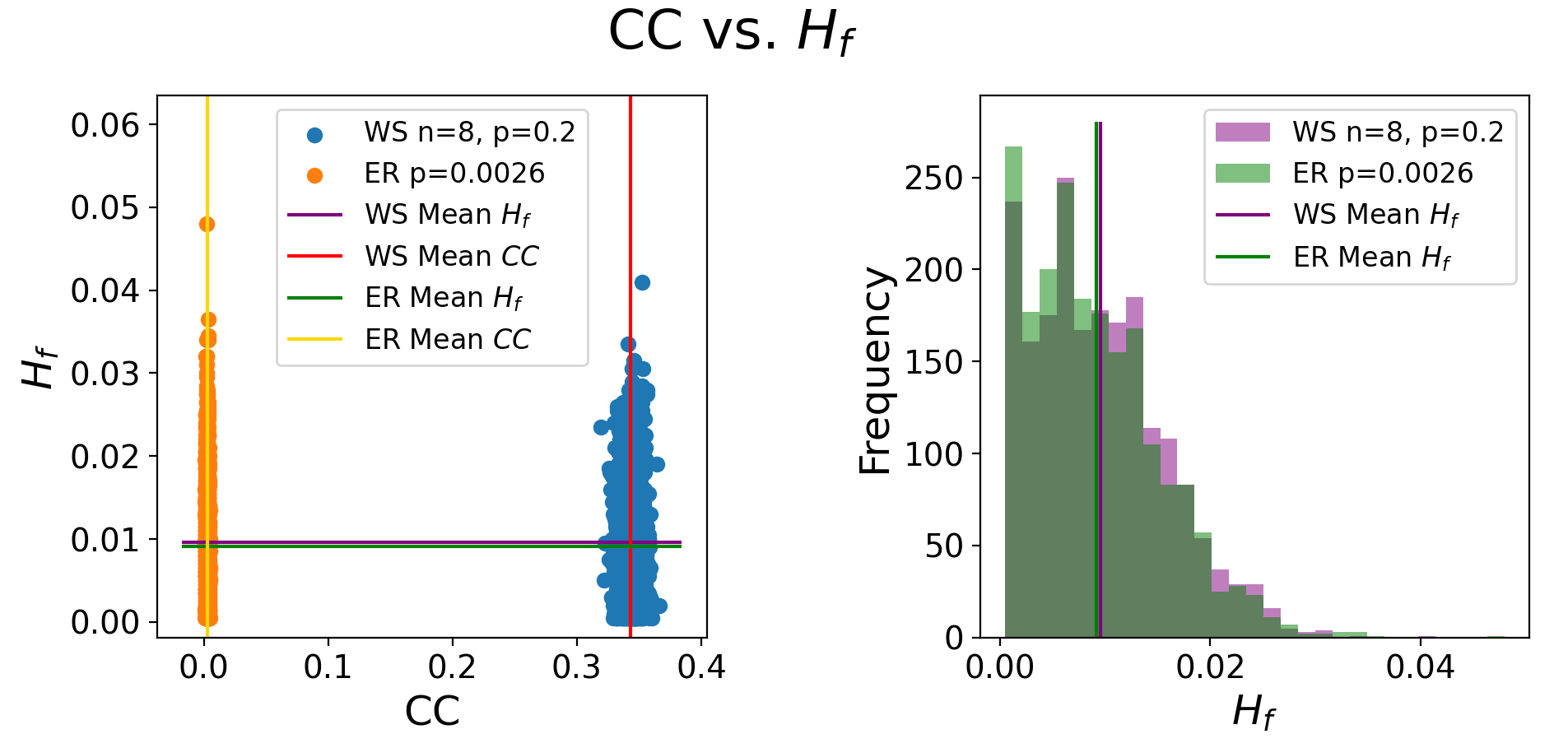}
    \vspace{-0.8cm}
    \captionof{figure}{Comparing the effects of the Erd\H{o}s-R\'{e}nyi network (p = 0.0026) and Watts-Strogatz (n = 8, p = 0.2) network on the ABM. On the left, clustering coefficient (CC) is plotted against the final proportion of $H$ individuals ($H_f$) for each realization. On the right, a plot is shown of the distribution of $H_f$ for each model realization. Observe that different CC can have very similar $H_f$ distributions, as further described by statistics in Tbl. \ref{tab:ks_es_erws}. This is evidence that CC is not a robust statistic for adequately predicting $H_f$ values.}
    \label{fig:ws_er_comp_CC}
\end{minipage}

\clearpage

\section{Parameter choices for all models used in optimization}
\label{sec:opt_controls}


A variety of network parameters for each given network generation algorithm were simulated with a resulting range of APL values. A comprehensive list of parameter choices is given below. Relationships between these parameters and their resulting APL values and $\beta_A,\beta_P,$ and $\theta_1$ parameters are shown in the following figures and tables.
\begin{itemize}
    \item \textbf{Erd\H{o}s-R\'{e}nyi}: $p$ =  \{0.00075, 0.0008, 0.001, 0.001125, 0.0015, 0.00175, 0.002, 0.0025, 0.003, 0.004, 0.005, 0.0075, 0.01, 0.025, 0.03, 0.05\}
    \item \textbf{Barab\'{a}si-Albert}: $m$ = \{2, 3, 4, 5, 6, 7, 8, 9, 11, 14, 17, 20\}
    \item \textbf{Watts-Strogatz}: $(n, p)$ = \{(2, 0.4), (2, 0.6), (2, 0.8), (2, 0.9), (4, 0.025), (4, 0.05), (4, 0.075), (4, 0.1), (4, 0.2), (4, 0.4), (4, 0.6), (4, 0.8), (6, 0.2), (6, 0.4), (6, 0.8), (8, 0.2), (8, 0.4), (8, 0.6), (8, 0.8), (10, 0.8), (10, 0.9), (12, 0.8)\}
\end{itemize}

\begin{figure}[h]
    \centering
    \includegraphics[width=\linewidth]{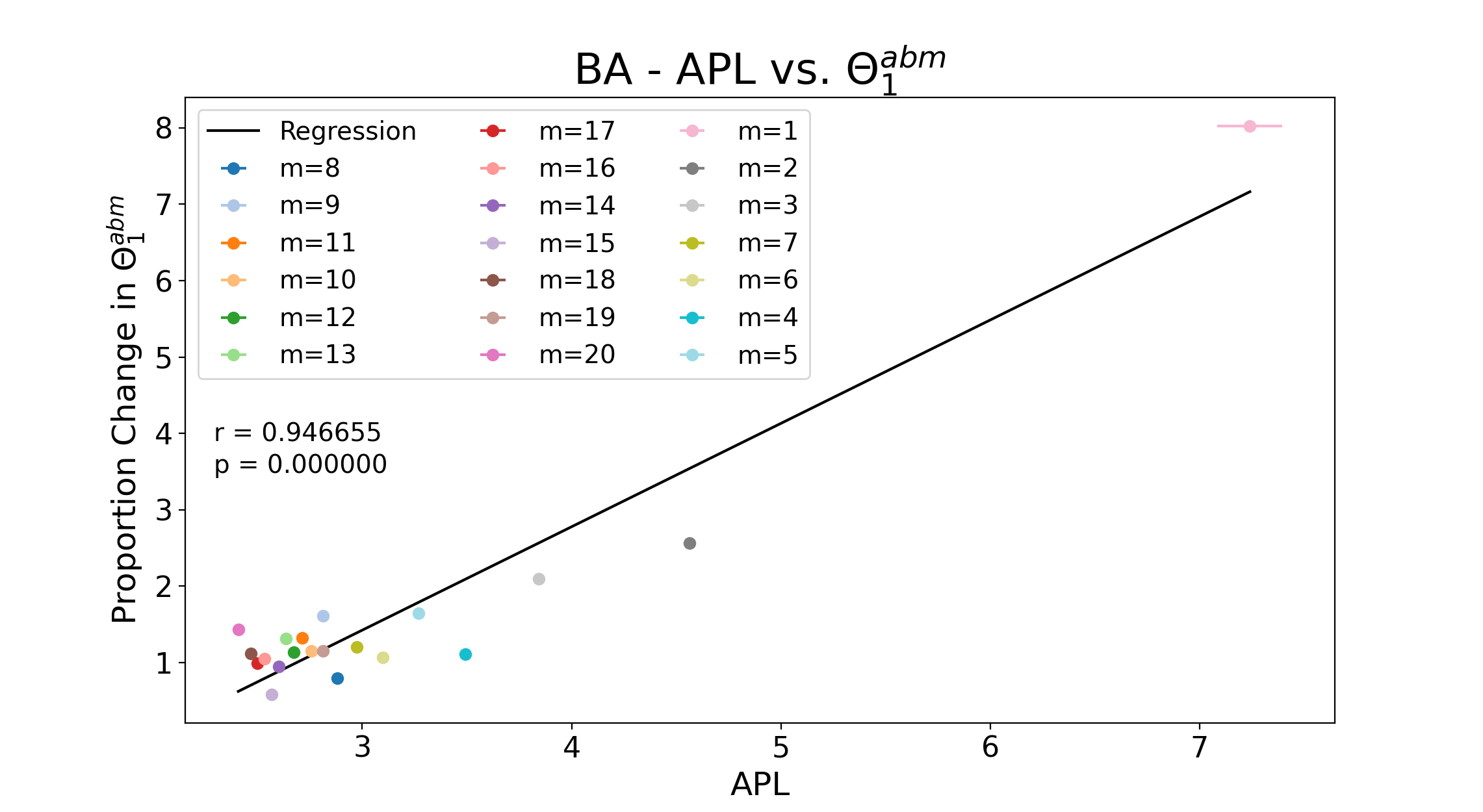}
    \caption{APL vs. inverted $\theta_1$ values for optimizations run on BA models with varying $p$ values. Each point represents the average APL for the derived $\theta_1$ value over 300 model runs, with each network in these 300 runs being generated with the same $p$ parameter. Horizontal bars are shown around each point corresponding to a 95\% confidence interval for the APL value. The Pearson's correlation coefficient ($r$) and $p$-value for the line of best fit are displayed.}
    \label{fig:ba_apl_v_theta}
\end{figure}

\begin{table}[h]
    \centering
    \footnotesize
    \begin{tabular}{ |p{1.3cm} p{1.5cm} p{1.6cm} p{1.6cm} p{1.6cm} p{2cm}| }
    \hline
    m & APL & $\beta_{A}^{abm}$ & $\beta_{P}^{abm}$ & $\theta_{1}^{abm}$ & final loss \\ 
    \hline
    2.0 & 4.5617 & 0.000878 & 6.54e-05 & 0.79145 & 5.66e-07 \\ 
    3.0 & 3.8435 & 0.000878 & 6.54e-05 & 0.68751 & 7.61e-07 \\ 
    4.0 & 3.4937 & 0.000878 & 6.54e-05 & 0.46757 & 8.22e-07 \\ 
    5.0 & 3.2689 & 0.000878 & 6.54e-05 & 0.58787 & 8.25e-07 \\
    \hline
    6.0 & 3.0979 & 0.000878 & 6.54e-05 & 0.45815 & 7.83e-07 \\ 
    7.0 & 2.9749 & 0.000878 & 6.54e-05 & 0.48997 & 8.59e-07 \\ 
    8.0 & 2.8815 & 0.000878 & 6.54e-05 & 0.3984 & 7.93e-07 \\ 
    9.0 & 2.8111 & 0.000878 & 6.54e-05 & 0.58072 & 7.24e-07 \\ 
    \hline
    10.0 & 2.7554 & 0.000878 & 6.54e-05 & 0.47798 & 6e-07 \\ 
    11.0 & 2.7124 & 0.000878 & 6.54e-05 & 0.51619 & 7.11e-07 \\ 
    12.0 & 2.6726 & 0.000878 & 6.54e-05 & 0.47346 & 8.01e-07 \\ 
    13.0 & 2.6347 & 0.000878 & 6.54e-05 & 0.51344 & 5.41e-07 \\ 
    \hline
    14.0 & 2.6005 & 0.000878 & 6.54e-05 & 0.43297 & 7.1e-07 \\ 
    15.0 & 2.5665 & 0.000878 & 6.54e-05 & 0.35185 & 1.23e-06 \\ 
    16.0 & 2.5316 & 0.000878 & 6.54e-05 & 0.45486 & 8.02e-07 \\ 
    17.0 & 2.4495 & 0.000878 & 6.54e-05 & 0.4418 & 7.25e-07 \\ 
    \hline
    18.0 & 2.4677 & 0.000878 & 6.54e-05 & 0.46984 & 7.53e-07 \\ 
    19.0 & 2.8111 & 0.000878 & 6.54e-05 & 0.47784 & 7.46e-07 \\ 
    20.0 & 2.4079 & 0.000878 & 6.54e-05 & 0.53986 & 7.31e-07 \\ 
    \hline
    \end{tabular}
    \caption{Optimization results for a variety of BA models. All parameters shown are inverted as previously described. The final loss is the mean squared error between the ABM projection mean and the ODE model projection for every class. Note that all values for $\beta_{A}^{abm}$ and $\beta_{P}^{abm}$ are constant across various APL values. These relationships are visualized in Fig. \ref{fig:ba_apl_v_theta}.}
    \label{tab:ba_opt_table}
\end{table}

\begin{figure}[h]
    \centering
    \includegraphics[width=\linewidth]{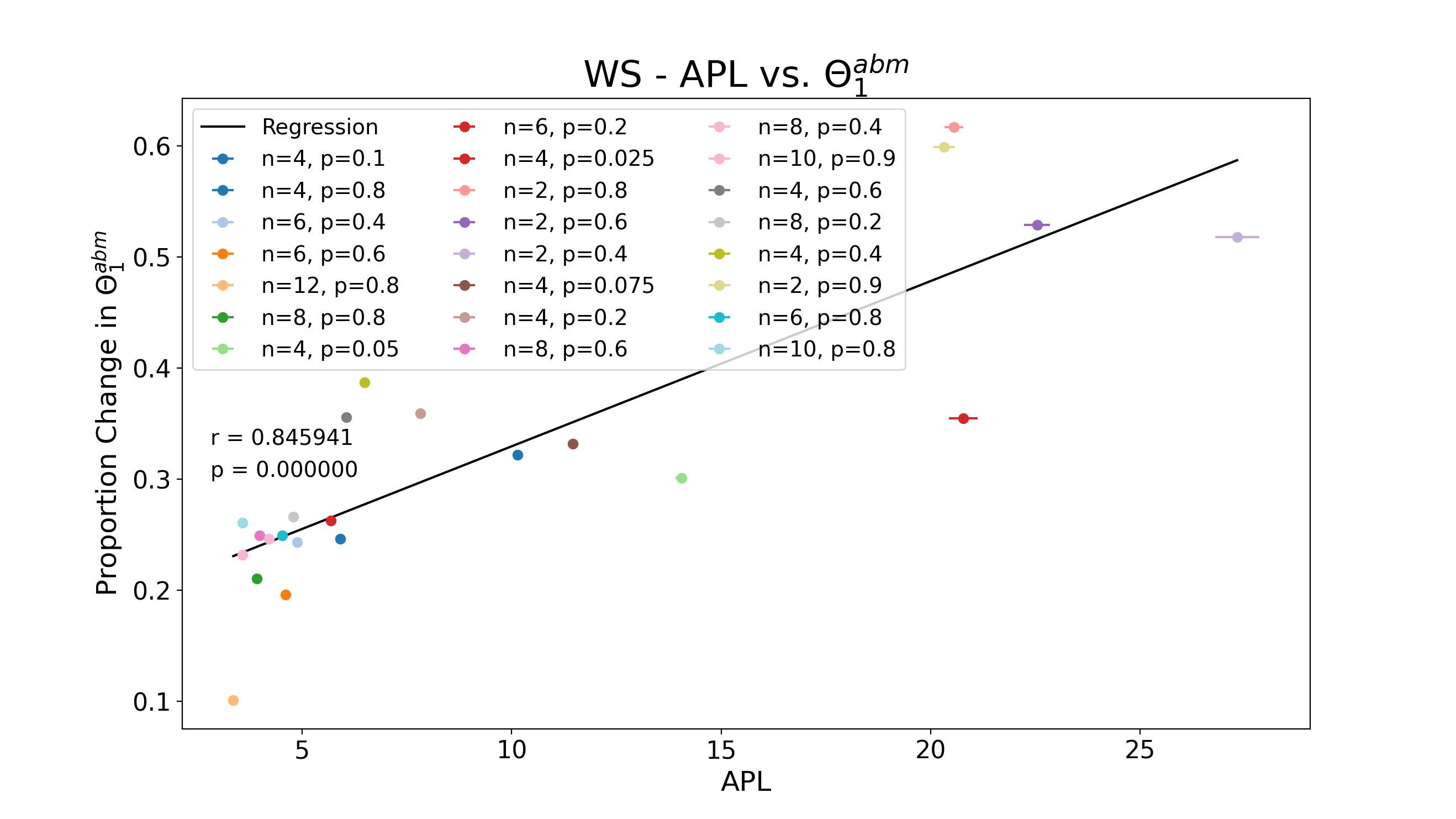}
    \caption{APL vs. inverted $\theta_1$ values for optimizations run on WS models with varying $p$ values. Each point represents the average APL for the derived $\theta_1$ value over 300 model runs, with each network in these 300 runs being generated with the same $p$ parameter. Horizontal bars are shown around each point corresponding to a 95\% confidence interval for that APL value. The Pearson's correlation coefficient ($r$) and the $p$-value for that $r$ calculation is also displayed.}
    \label{fig:ws_apl_v_theta}
\end{figure}

\begin{table}[h]
    \centering
    \footnotesize
    \begin{tabular}{ |p{0.9cm} p{1.2cm} p{1.4cm} p{1.8cm} p{1.6cm} p{1.6cm} p{2cm}| }
    \hline
    n & p & APL & $\beta_{A}^{abm}$ & $\beta_{P}^{abm}$ & $\theta_{1}^{abm}$ & final loss \\ 
    \hline
    4 & 0.8 & 5.9044 & 0.000878 & 6.54e-05 & 0.27668 & 5.3e-07 \\ 
    6 & 0.4 & 4.8804 & 0.000878 & 6.54e-05 & 0.27598 & 5.22e-07 \\ 
    6 & 0.6 & 4.6079 & 0.000878 & 6.54e-05 & 0.26558 & 4.04e-07 \\ 
    12 & 0.8 & 3.358 & 0.000878 & 6.54e-05 & 0.24442 & 5.54e-07 \\ 
    \hline
    8 & 0.8 & 3.9206 & 0.000878 & 6.54e-05 & 0.26876 & 6.39e-07 \\ 
    4 & 0.05 & 14.053 & 0.000878 & 6.54e-05 & 0.28881 & 4.22e-07 \\ 
    6 & 0.2 & 5.6838 & 0.000878 & 6.54e-05 & 0.28033 & 3.89e-07 \\ 
    4 & 0.025 & 20.783 & 0.000878 & 6.54e-05 & 0.30081 & 4.44e-07 \\
    \hline
    2 & 0.8 & 20.553 & 0.000878 & 6.54e-05 & 0.35899 & 4.45e-07 \\ 
    2 & 0.6 & 22.541 & 0.000878 & 6.54e-05 & 0.33946 & 3.87e-07 \\ 
    2 & 0.4 & 27.313 & 0.000878 & 6.54e-05 & 0.33697 & 3.9e-07 \\ 
    4 & 0.075 & 11.465 & 0.000878 & 6.54e-05 & 0.2957 & 4.95e-07 \\ 
    \hline
    4 & 0.2 & 7.8187 & 0.000878 & 6.54e-05 & 0.30175 & 6.45e-07 \\ 
    8 & 0.6 & 3.9917 & 0.000878 & 6.54e-05 & 0.27736 & 4.95e-07 \\ 
    8 & 0.4 & 4.2069 & 0.000878 & 6.54e-05 & 0.27673 & 2.59e-07 \\ 
    10 & 0.9 & 3.5739 & 0.000878 & 6.54e-05 & 0.27353 & 3.21e-07 \\
    \hline
    4 & 0.6 & 6.0563 & 0.000878 & 6.54e-05 & 0.30102 & 3.95e-07 \\ 
    8 & 0.2 & 4.792 & 0.000878 & 6.54e-05 & 0.28115 & 3.32e-07 \\ 
    4 & 0.4 & 6.488 & 0.000878 & 6.54e-05 & 0.30793 & 3.8e-07 \\ 
    2 & 0.9 & 20.323 & 0.000878 & 6.54e-05 & 0.35503 & 4.34e-07 \\ 
    \hline
    6 & 0.8 & 4.5212 & 0.000878 & 6.54e-05 & 0.27735 & 3.32e-07 \\ 
    10 & 0.8 & 3.5795 & 0.000878 & 6.54e-05 & 0.27986 & 4.27e-07 \\ 
    \hline
    \end{tabular}
    \caption{Optimization results for a variety of WS models. All parameters shown are inverted as previously described. The final loss is the mean squared error between the ABM projection mean and the ODE model projection for every class. Note that all values for $\beta_{A}^{abm}$ and $\beta_{P}^{abm}$ are constant across various APL values. These relationships are visualized in Fig. \ref{fig:ws_apl_v_theta}.}
    \label{tab:ws_opt_table}
\end{table}

\end{document}